\documentclass[twocolumn,twocolappendix,numberedappendix,appendixfloats,fleqn]{openjournal}

\usepackage{float}
\usepackage{graphicx,amsmath,amssymb,amstext}
\usepackage{graphicx}
\usepackage{url}
\usepackage{natbib}
\usepackage{orcidlink}
\usepackage{hyperref}
\hypersetup{
    colorlinks=true,
    linkcolor=blue,
    filecolor=blue,      
    urlcolor=blue,
    citecolor=blue,
}

\newcommand*    {\diff}     {\mathop{}\!\mathrm{d}}

\begin{document}

\title{Power Spectra of JWST images of Local Galaxies: Searching for Disk Thickness}

\author{Bruce~G.~Elmegreen\,\orcidlink{0000-0002-1723-6330}$^{1}$}
\author{Angela~Adamo\,\orcidlink{0000-0002-8192-8091}$^{2}$}
\author{Varun~Bajaj\,\orcidlink{0009-0008-4009-3391}$^{3}$}
\author{Ana~Duarte-Cabral\,\orcidlink{0000-0002-5259-4774}$^{4}$}
\author{Daniela Calzetti\,\orcidlink{0000-0002-5189-8004}$^5$}
\author{Michele Cignoni\,\orcidlink{0000-0001-6291-6813}$^{6,7,8}$}
\author{Matteo Correnti\,\orcidlink{0000-0001-6464-3257}$^{9,10}$}
\author{John S. Gallagher, III\,\orcidlink{0000-0001-8608-0408}$^{11}$}
\author{Kathryn Grasha$^*$\,\orcidlink{0000-0002-3247-5321}$^{12,13}$}
\author{Benjamin Gregg\,\orcidlink{0000-0003-4910-8939}$^5$}
\author{Kelsey E. Johnson\,\orcidlink{0000-0001-8348-2671}$^{14}$}
\author{Sean T. Linden\,\orcidlink{0000-0002-1000-6081}$^{15}$}
\author{Matteo Messa\,\orcidlink{0000-0003-1427-2456}$^6$}
\author{G\"oran \"Ostlin\,\orcidlink{0000-0002-3005-1349}$^2$}
\author{Alex Pedrini\,\orcidlink{0000-0002-8222-8986}$^2$}
\author{Jenna Ryon\,\orcidlink{0000-0002-2918-7417}$^3$}

\affiliation{$^1$Katonah, NY 10536, USA}
\affiliation{$^2$Department of Astronomy, The Oskar Klein Centre, Stockholm University, AlbaNova, SE-10691 Stockholm, Sweden}
\affiliation{$^3$Space Telescope Science Institute, 3700 San Martin Drive Baltimore, MD 21218, USA}
\affiliation{$^4$School of Physics \& Astronomy, Cardiff University, Queen’s Building, The Parade, Cardiff CF24 3AA, UK}
\affiliation{$^5$Department of Astronomy, University of Massachusetts Amherst, 710 North Pleasant Street, Amherst, MA 01003, USA}
\affiliation{$^6$INAF—Osservatorio di Astrofisica e Scienza dello Spazio di Bologna, Via Gobetti 93/3, I-40129 Bologna, Italy}
\affiliation{$^7$Department of Physics, University of Pisa, Largo B. Pontecorvo 3, 56127 Pisa, Italy}
\affiliation{$^8$INFN, Largo B. Pontecorvo 3, 56127 Pisa, Italy}
\affiliation{$^9$INAF Osservatorio Astronomico di Roma, Via Frascati 33, 00078 Monteporzio Catone, Rome, Italy}
\affiliation{$^{10}$ASI-Space Science Data Center, Via del Politecnico, I-00133 Rome, Italy}
\affiliation{$^{11}$Department of Physics and Astronomy, Macalester University, 1600 Grand Avenue, Saint Paul, MN 55105-1899, USA}
\affiliation{$^{12}$Research School of Astronomy and Astrophysics, Australian National University, Canberra, ACT 2611, Australia}
\affiliation{$^{13}$ARC Centre of Excellence for All Sky Astrophysics in 3 Dimensions (ASTRO 3D), Australia}
\affiliation{$^{14}$Department of Astronomy, University of Virginia, Charlottesville, VA, USA}
\affiliation{$^{15}$Steward Observatory, University of Arizona, 933 N Cherry Avenue, Tucson, AZ 85721, USA}

\thanks{$^*$ARC DECRA Fellow}

\begin{abstract}
JWST/MIRI images have been used to study the Fourier transform power spectra (PS) of two spiral galaxies, NGC 628 and NGC 5236, and two dwarfs, NGC 4449 and NGC 5068, at distances ranging from 4 to 10 Mpc. The PS slopes on scales larger than 200 pc range from $-0.6$ at $21\mu$m to $-1.2$ at $5.6\mu$m. These slopes for one-dimensional PS are consistent with the PS slopes observed elsewhere using HI and dust emission. They are likely related to turbulence, but they may also be viewed as a hierarchical distribution of objects having a size-luminosity relation and size distribution function. There is no evidence for a kink or steepening of the PS at some transition from two-dimensional to three-dimensional turbulence on the scale of the disk thickness. This lack of a kink could be from large positional variations in the PS depending on two opposite effects: local bright sources that make the slope shallower and exponential galaxy profiles that make the slope steeper.  The sources could also be confined to a layer of molecular clouds that is thinner than the HI or cool dust layers where PS kinks have been observed before.  If the star formation layers observed in the mid-infrared here are too thin, then the PS kink could also be hidden in the broad tail of the JWST point spread function.
\end{abstract}

\section{Introduction} \label{sec-intro}
\begin{figure}
    \includegraphics[width=\columnwidth]{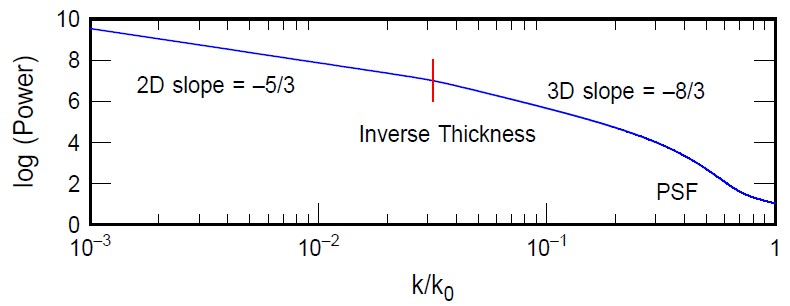}
    \caption{Schematic of the PS of a galaxy disk with a shallow slope at small wavenumber $k$, from 2D turbulence, representing large scales, and a larger slope at large $k$ from 3D turbulence. The dip at high $k$ from the point spread function is also shown. Equations 2 and 3 in \cite{koch20} were used.}
    \label{fig:schematic}
\end{figure}

The power spectrum (PS) of interstellar gas emission reveals the relative importance of structures and their luminosities over a wide range of scales, as measured by the Fourier transform wavenumber, $k$, which is the inverse of length. An important signature observed in the LMC \citep{elmegreen01,block10}, M33 \citep{combes12}, and the dwarf galaxy NGC 1058 \citep{dutta09a} is a steepening of the PS at wavenumbers exceeding the inverse disk thickness, as expained theoretically by \cite[][eq. 28]{lazarian00}. This steepening reflects a transition from two to three dimensional turbulence. Numerical simulations confirm this signature in both the PS and spectral correlation function \citep{padoan01,bournaud10,combes12, fensch23,tress20}. A schematic of the PS kink is shown in Fig.~\ref{fig:schematic}. 

PS observations of HI emission by \cite{szot19} suggest that the LMC thickness ranges from $\sim50$ pc near 30 Dor to $\sim250$ pc in the central region and $\sim400$ pc in the outer region.  The two-dimensional (2D) PS slope at small $k$ ranges from approximately $-1.5$ in the central regions to $-3.5$ in the outer regions, while the high $k$ slope ranges from $-3$ to $-4.5$ in these regions, respectively.  The PS break in M33 observed by \cite{combes12} depends on wavelength, ranging from $\sim50$ pc in the ultraviolet to $\sim100$ pc in HI and CO, and $\sim300$ pc in H$\alpha$, with a similar range in the infrared. For most of these wavelengths, the 2D PS slope is $\sim-1.5$ at low $k$ and $\sim-3.5$ at large $k$. For NGC 1058, \cite{dutta09a} observed the PS kink in HI at $490\pm90$ pc, with 2D PS slopes of $-1.0$ at low $k$ and $-2.5$ at high $k$.

With such a signature in the PS, the thicknesses of face-on galaxies can be inferred and mapped.  Thickness can be important in converting gas surface density to volume density for determination of the local gas free-fall time \citep{bacchini20}, or for correcting an observed rotation curve for radial pressure gradients in order to assess the distribution and amount of dark matter \citep{verbeke17}.

The explanation for the change in PS slope at the size scale of the disk thickness was illustrated in \cite{bournaud10} and \cite{fensch23} as the result of three components of turbulent velocity producing structure on small scales and two components of velocity producing structure on large scales. Turbulent energy is also dispersed into a larger volume of phase space in 3D than in 2D, making the energy cascade steeper in 3D. 

To see the disk thickness in a PS, the spatial resolution of the galaxy has to exceed the disk thickness by an order of magnitude or more. Then the slope of the steeper PS at high $k$ can be distinguished from the slope of the shallower PS at low $k$. The ratio of the galaxy size to the thickness should also exceed a factor of $\sim10$, so there is another order of magnitude at small $k$, i.e., on lengths larger than the thickness. These two constraints are somewhat incompatible for a given angular resolution, because galaxies close enough to resolve for the first constraint tend to be dwarfs, which are also typically too thick \citep{bergh88} to satisfy the second constraint.

A third requirement to see disk thickness in a PS is that the emission should come from sources that are related to correlated motions such as interstellar turbulence, and not thermal motions as in warm or hot gas. Random stellar motions should not show the disk thickness in a PS.  Thus the PS of light from old stars would not have a kink at the disk thickness, but the PS of star-forming regions traced by young stars could have it because of the turbulent structure of the gas that forms these stars. Thickness signatures in the PS of spiral galaxies observed at high resolution in the B-band are an example \citep{elmegreen03a,elmegreen03b}.  On the other hand, there were no significant kinks in the PS of 9 dwarfs studied at V band by \cite{willett05}, presumably because these galaxies are relatively thick on the line of sight.  The H$\alpha$ images in \cite{willett05} had power-law PS with slopes that decreased from near-zero (characteristic of noise) to the same negative values as the V-band images, which was  $-1.4\pm0.3$ for one-dimensional scans, as the H$\alpha$ filling factor increased from $\sim0$ to $\sim1$ with the star formation rate. 

Here we investigate the PS of JWST images of dust emission from nearby galaxies at $5.6\mu$m to $21\mu$m wavelength. We use data on NGC 628, NGC 4449, and NGC 5236 from the  “Feedback in Emerging extrAgalactic Star clusTers” (FEAST) survey, and additional data for  NGC 628 and a nearby galaxy, NGC 5068, from the PHANGS-JWST survey \citep[GO 2107,][]{lee23}.  
These galaxies have the potential to satisfy all three of the above requirements to measure the disk thickness because of the high angular resolution of JWST and the small distances. However, problems can arise if the broad wings of the Point Spread Function (PSF) blend with the thickness scale, or if the closest galaxies are intrinsically thick.

Two of our galaxies were studied before. \cite{dutta08} measured the 2D PS of HI emission from NGC 628, finding a continuous slope of $-1.6\pm0.2$ from 8 kpc to 800 pc, which was too large a scale to observe a possible thickness effect at $\sim200$ pc. \cite{grisdale17} also observed NGC 628 in HI; they found a slope of $-2.2$ with no kink down to 400 pc. For NGC 5236, \cite{nandakumar20} measured a continuous 2D PS slope of $-1.23\pm0.06$ from 11 kpc to 300 pc for HI column density. 

The PS has also been determined for several other galaxies at various wavelengths. In the first galaxy-scale measurement, \cite{stan99} observed HI emission from the Small Magellanic Cloud (SMC) and found a continuous 2D PS with a slope of $-3.04\pm0.02$ in narrow velocity intervals over spatial scales from 30 pc to 4 kpc \citep[see also][]{pingel22}.  The slope steepens with increasing velocity width, allowing a separation of the PS for velocity and density \citep{lazarian00,dutta15}. The density alone had a PS slope of $-3.3\pm0.01$ from the integrated velocity map \citep{stanimirovic01}, which is about the same as that obtained from a dust column density map of the SMC \citep{stan00}.  In the Magellanic bridge, \cite{muller04} measured a power-law PS in the southwest region over scales from 29 pc to $2$ kpc, finding a slope equal to $-3.02\pm0.14$ for narrow velocity channels and steepening to $-3.4$ for the full velocity width. The northern region of the Magellanic bridge has a shallower slope. 

In the dwarf DDO 210, \cite{begum06} found a 2D HI PS slope of $-2.75\pm0.45$ from 80 pc to 500 pc. They suggested that the similarity of this slope to the slopes in the LMC, SMC and Milky Way implies that interstellar structure is independent of the star formation rate. \cite{dutta09b} also reported the 2D HI PS for seven dwarfs and one spiral, obtaining slopes ranging from $-1.5$ in And IV, NGC 628, UGC 4459, and GR 8 to $-2.6$ in DDO 210 and NGC 3741.  They interpreted the shallow slopes as the result of 2D turbulence in galaxies where the thickness could not be resolved, and the steep slopes as the result of 3D turbulence everywhere for intrinsically thick galaxies.  They noted that the shallow slope correlates with the star formation rate per unit area, with larger rates corresponding to steeper slopes. Such a correlation between star formation rate and slope was also noted for various regions around the LMC by \cite{szot19}, who suggested that star formation feedback destroys small scale structure in HI. 

\cite{zhang12} determined 2D PS of HI for 24 dwarf irregular galaxies over single velocity channels and over  the integrated velocities. The length scales typically ranged between several hundred pc and several kpc but in some cases were smaller. There were no kinks suggesting thickness for these galaxies, but the larger galaxies systematically had shallower slopes, as if they were everywhere thin on the line of sight compared to the smaller galaxies, which could be thicker. The division between these cases occurred at a star formation rate of $10^{-2.1}\;M_\odot$ yr$^{-1}$ and an absolute magnitude of $M_B=-14.5$ mag. The slopes also varied from a constant value at large velocity widths to shallower values at narrow velocity widths, which suggested relatively more small-scale structure beyond the limit of velocity resolution resulting from cool neutral gas, particularly in the inner galactic regions. \cite{zhang12} noted that although the slopes depended on absolute SFR as a result of galaxy size, they were independent of the average SFR surface density, which led them to suggest that non-stellar power sources drive the implied turbulence. A similar conclusion was reached by \cite{nest17} for the SMC.

Spiral galaxies have also had PS measurements in HI, but usually without the resolution necessary to see a feature from disk thickness. \cite{dutta10} calculated the HI PS for NGC 4254, finding a slope of $-1.7$ on scales ranging from 1.7 kpc to 8.4 kpc. They suggested NGC 4254 is a harassed galaxy, and noted that the PS slope in the outer parts, $-2.0\pm0.3$, is steeper than in the inner parts, $-1.5\pm0.2$. \cite{dutta13a} measured the HI PS slopes for 18 spiral galaxies over scales ranging from several hundred pc to 16 kpc; the PS were power laws with slopes of around $-1.3$, which they concluded was from 2D turbulence. \cite{dutta13b} then showed how to convert the amplitude of the PS to the HI mass fraction. More recently, \cite{nandakumar23} measured a PS slope of $-0.96\pm0.05$ for HI column density and $-1.81\pm0.07$ for velocity in NGC 6946. They suggested a driving scale larger than 6 kpc, with magnetic processes contributing to solenoidal motions and gravitational processes contributing to compressive motions.

\cite{grisdale17}  measured PS of the surface densities for six THINGS galaxies, finding large-scale slopes of $-2.2$ for NGC 628 (as noted above), $-2.8$ for NGC 3521, $-2.1$ for NGC 4736, $-2.2$ for NGC 5055, $-2.5$ for NGC 5457, and $-1.6$ for NGC 6946. No slope changes were seen down to scales of several hundred pc that might indicate the scale of the disk thickness. 

Most of the PS slopes discussed above were determined for two-dimensional surfaces, such as whole galaxies or regions of galaxies, and are steeper by 1 than the slopes derived here, which are for one-dimensional intensity strips. Strips have the advantage that they can avoid excessively bright sources, which are a problem for near-infrared images because of star formation. Bright sources flatten the PS and can distort the average over a disk region (see below). We also consider azimuthal intensity scans for NGC 628 to avoid a component of the PS from the exponential radial profile of the disk, which is clearly seen here in minor or major axis strips near the galaxy centers. Azimuthal profiles also put the spiral arms in the lowest few bins of wavenumber so the spiral is easily isolated; i.e., the two-arm component of the spiral has $k=2$ in a scan that might go up to $k=10^3$ or more.  A third advantage of azimuthal strips is that the intensity is a periodic function of the scan position, making it ideal for Fourier transforms.  A second paper (Elmegreen et al. 2025, in preparation) measures the PS from azimuthal intensity scans of M51.

The PS kink as a tracer of disk thickness has been questioned by \cite{koch20}, who looked at FIR emission maps and dust surface densities maps of the LMC, SMC, M31 and M33. The most important test was for the LMC because the SMC has no kink in HI and the spatial resolutions for M31 and M33 were comparable to their expected thicknesses. For the LMC, \cite{koch20} suggest the PS is a continuous power law outside of the 30 Dor region, and the kink in the whole galaxy PS comes from 30 Dor.  \cite{szot19} map the PS slopes over the LMC and also get a steeper PS near 30 Dor, but they also find PS kinks elsewhere in the LMC. 

In summary, the previous observations rarely detected a disk thickness from two components of the PS. Usually the PS had one continuous power law from scales of multiple kpc down to the resolution limit, which was typically larger than the expected thickness of several hundred pc. The continuous power law still gave important information, however, confirming expectations from 2D turbulence theory and pointing to very large driving scales at the lowest wavenumbers. This implied there are large-scale energy sources and continuous energy cascades or structural hierarchies below that.

In what follows, Section \ref{sect:data}  summarizes the data used here. Section \ref{sect:628_560_intensity} shows sample intensity scans for NGC 628 at $5.6\mu$, Section \ref{sect:628_560_PS} presents the PS for these scans, and Section \ref{sect:628_770_1000_2100_PS} gives comparable results for NGC 628 at $7.7\mu$m, $10\mu$m and $21\mu$m.  Section \ref{sect:5236_4449_5068} then considers the other three galaxies: NGC 5236 and NGC 4449 at $5.6\mu$m and $7.7\mu$m, and NGC 5068 at $10\mu$m at $21\mu$m. In Section \ref{sect:lowI}, we recalculate all of these PS using only the scans with the lowest peak intensities in order to isolate what may be a disk component between the star-forming regions. We do the same type of intensity thresholding in Section \ref{sect:azimuthal}, using azimuthal scans of NGC 628. 
The results are discussed in Section \ref{sect:discussion} and the conclusions are in Section \ref{sect:conclusions}.
An Appendix has additional figures.

\section{FEAST Survey}
\label{sect:data}
The FEAST survey is a cycle-1 JWST program (GO 1783, PI Adamo) designed to map the early stages of star cluster formation in six representative local galaxies. JWST observations with NIRCam and MIRI cameras sample the continuum, H recombination lines (Pa$\alpha$ and Br$\alpha$), and three bands from 1 to 8 $\mu$m (Adamo et al in prep.). In the present paper, we use the MIRI observations in F560W and F770W filters for NGC 628, NGC 4449, and NGC 5236, and MIRI data also in F1000W and F2100W for NGC 628 from PHANGS-JWST \citep{lee23}.
Relevant parameters for all the galaxies considered here are in Table 1. 

The MIRI observations consist of mosaics of 1$\times$5 or $1\times3$ tiles and are completed with external background observations. Stage-two pipeline frames of the background were used to create a master background for each of the MIRI filters, and then subtracted from single target exposures during level-two processing. We conducted the sky matching step between different exposures using {\it PixelSkyMatchStep} \citep{VarunSoftware}. The remainder of the level-three processing created a single mosaic for each filter. The final MIRI mosaics are resampled to a scale of 0.08\arcsec/px.

We also include in this analysis archival MIRI observations of NGC 5068 from the PHANGS-JWST cycle 1 large program, available at $0.11^{\prime\prime}$ pixel size \citep{lee23}. The mosaics consisted of 1$\times$3 tiles and 1 external background observation. Data acquisition and reduction are described in \cite{williams24} and data are accessible at the PHANGS repository\footnote{\url{https://archive.stsci.edu/hlsp/phangs/phangs-jwst}}.

\begin{deluxetable}{ccccccc}
\tablecaption{Galaxy Sample and Scan Details \label{table:sample}}
\tablehead{\colhead{Galaxy}
&\colhead{Dist.}
&\colhead{Filter}
&\colhead{Pixel}
&\colhead{FWHM}
&\colhead{Length}
&\colhead{Number}
\\
NGC&Mpc&&${\prime\prime}$
&pc$^a$&pixels&of scans}
\startdata
628  & 9.84 & F560W  & 0.08 & 9.9  & 1408 & 4055 \\
628  & 9.84 & F770W  & 0.08 & 12.8 & 1408 & 4055 \\
628  & 9.84 & F1000W & 0.08 & 15.6 & 1546 & 2734 \\
628  & 9.84 & F2100W & 0.08 & 32.1 & 1546 & 2734 \\
5236 & 4.66 & F560W  & 0.08 & 4.7 &  1472 & 4232 \\
5236 & 4.66 & F770W  & 0.08 & 6.1 &  1472 & 4232 \\
4449 & 4.27 & F560W  & 0.08 & 4.3 &  1517 & 2530 \\
4449 & 4.27 & F560W  & 0.08 & 4.3 &  2444$^b$ & 1517$^b$ \\
4449 & 4.27 & F770W  & 0.08 & 5.6 & 1516 & 2530 \\
4449 & 4.27 & F770W  & 0.08 & 5.6 & 2444$^b$ & 1517$^b$ \\
5068 & 5.2 & F1000W & 0.11 & 8.3 & 1380 & 1908  \\
5068 & 5.2 & F2100W & 0.11 & 17.0 & 1380 & 1908  \\
\enddata
\footnotetext{$^a$FWHM resolutions in pc, converted from the arcsec FWHM in the JWST User Documentation (see footnote 1), using the distances.}
\footnotetext{$^b$ Major axis intensity scan.}
\end{deluxetable}

\section{One-Dimensional Power Spectra of NGC 628 at $5.6$ microns}
\label{sect:628_560}
\subsection{Intensity Scans}
\label{sect:628_560_intensity}

\begin{figure}
    \begin{center}
        \includegraphics[width=6cm]{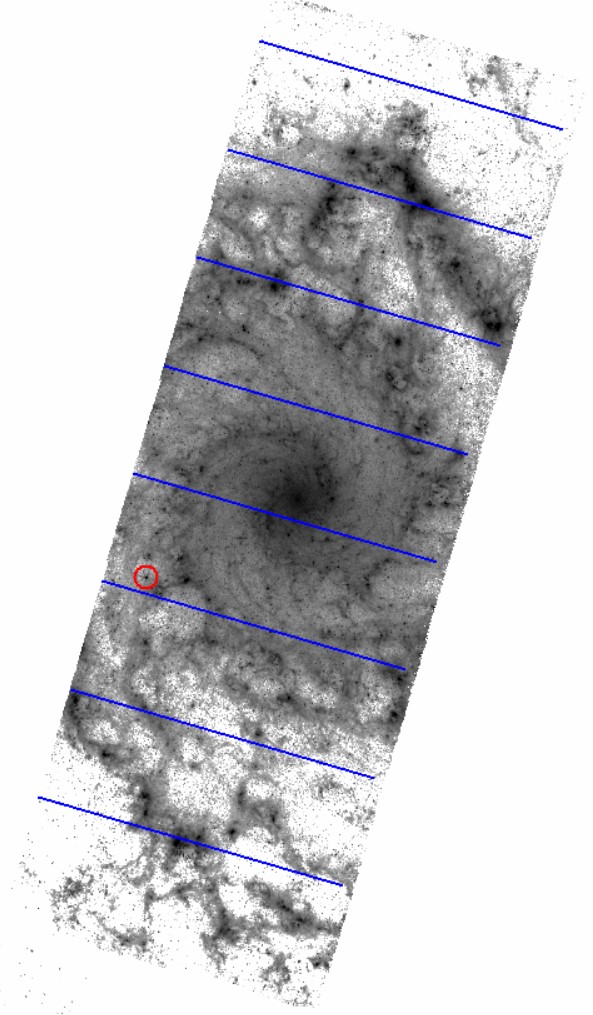}
    \end{center}
    \caption{NGC 628 at $5.6\mu$m with sample scan directions like those used to determine the PS. The separation between lines is 500 pixels, which is $40^{\prime\prime}$, or 1.91 kpc. The red circle shows the bright source used to show the PSF at $5.6\mu$m. 
    }
    \label{fig:628image}
\end{figure}

Fig.~\ref{fig:628image} shows the FEAST image of NGC 628 taken with the JWST F560W filter. The nominal Full Width at Half Maximum (FWHM) of the PSF is $0.207^{\prime\prime}$ from the JWST User Documentation\footnote{\url{https://jwst-docs.stsci.edu/jwst-mid-infrared-instrument/miri-performance/miri-point-spread-functions}}.
This FWHM corresponds to 2.59 pixels and 9.9 pc at the distance of 9.84 Mpc to NGC 628 \citep[from][]{anand21,leroy21,tully09}. Overlaid on the image are 8 lines showing representative locations of intensity scans used to determine one-dimensional PS. Each scan is 1408 pixels long, 1 pixel wide, and for the entire length of the image, there are 4055 such scans (scan parameters for all of the galaxies are in Table 1). The figure shows every 500th scan. Also shown is a red circle surrounding a small bright source that will be used to illustrate the PSF.

We use 1D intensity scans for the PS in order to get a long length but not cover a large area, which might include several types of regions. We expect the dust disk scale height to be in a range from 100 pc to 300 pc based on NIR and CO images of the edge-on galaxy NGC 891 \citep{scoville93,bocchio16,elmegreen20,chastenet24}. This scale corresponds to a range from 26 pixels to 79 pixels for NGC 628, so a scan 1408 pixels long will have a presumed break point somewhere close to the middle on a plot of the log of the wavenumber. That is, $\log 26$ and $\log 79$ are 0.45 and 0.60 times $\log 1408$, respectively. Thus the indicated scans are the right length to show a PS kink if there is one. Taking all $\sim4000$ scans along the length of the FEAST image effectively maps the PS and any thickness indicators over the face of the disk.  

The intensity along a scan was sampled with 1 pixel spacing, but the scan pixels are not exactly the same as the image pixels. To determine the intensity at the center of a scan pixel, we averaged over the values in the scan pixel and the nearest 3 other pixels,  using weights proportional to the pixel areas included within $\pm0.5$ pixel of the scan pixel. Each scan had the same length with endpoints that lied entirely within the image and no zeros at either end. The scans are perpendicular to the long axis of the FEAST image. The data have been cleaned in the reduction process so there are no bad pixels, which may be expected towards the edges of the detectors.

\begin{figure}
    \includegraphics[width=\columnwidth]{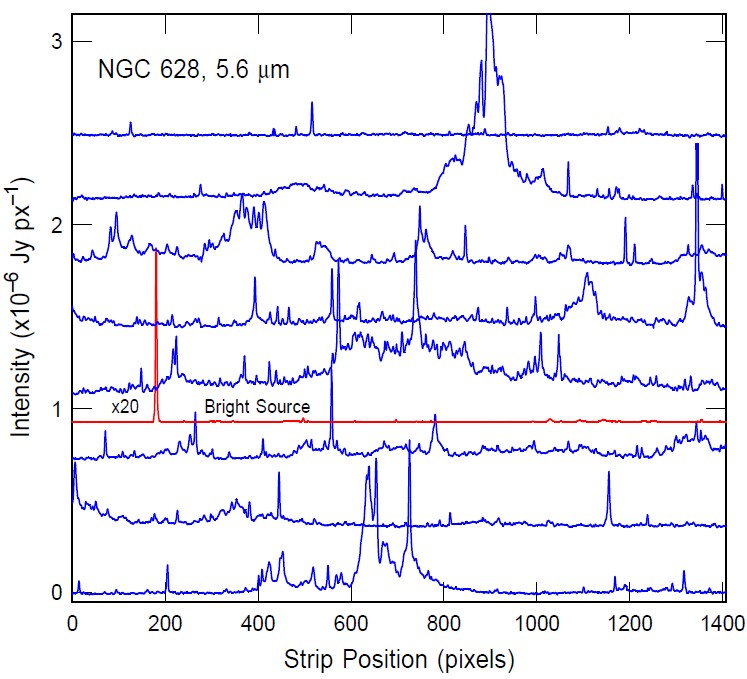}
    \caption{Intensity scans across the $5.6\mu$m image of NGC 628, shifted vertically from bottom (south) to top (north) for better viewing, that correspond to the blue lines in Fig.~\ref{fig:628image}. The red curve, plotted at 1/20th scale compared to the others, corresponds to the intensity scan through the bright $5.6\mu$m source indicated by the red circle in the previous image. }
    \label{fig:intensity}
\end{figure}

The blue curves in Fig.~\ref{fig:intensity} show intensity scans for the 8 equally-spaced positions corresponding to the lines in Fig.~\ref{fig:628image}. The red curve is the intensity scan through the bright source at  $1^{\rm h}36^{\rm m}45.46^{\rm s}$,  $15^\circ46^\prime33.43^{\prime\prime}$, which is shown by the red circle in Fig.~\ref{fig:628image}. The bright-source intensity has been divided by 20 to fit on the axes. 

\begin{figure}
    \hspace*{-4mm}\includegraphics[width=93mm]{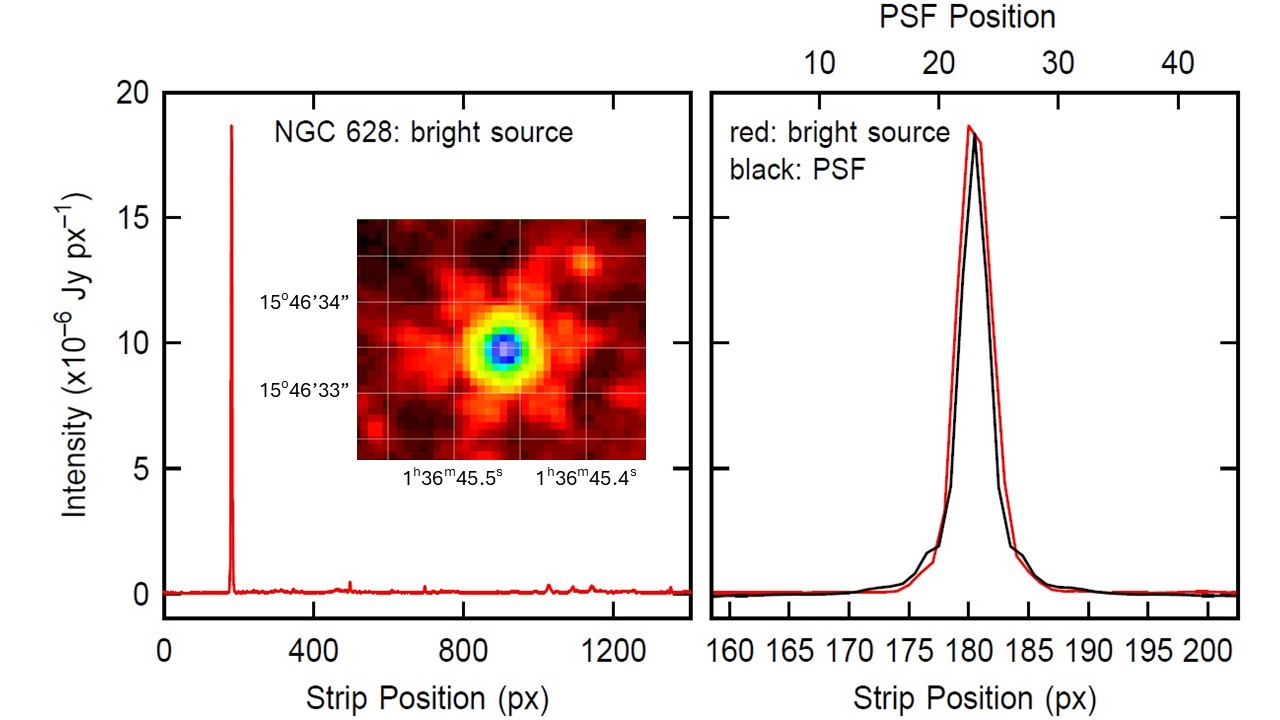}
    \caption{\textbf{Left}: The red curve is the same intensity scan as that shown by a red curve in Fig.~\ref{fig:intensity}. It is one pixel wide and goes through the small bright source indicated by the red circle in Fig.~\ref{fig:628image}. The insert shows the source itself; the coordinates are $1^{\rm h}36^{\rm m}45.46^{\rm s}$,  $15^\circ46^\prime33.43^{\rm s}$.
    \textbf{Right}: The red curve is a blow-up of the same scan highlighting the bright source, and the black curve is a scan through an idealized PSF made from bright point sources in the same image. The strip position on the lower abscissa is in reference to the bright-source scan through the galaxy, while the PSF position at the top is in reference to the idealized two-dimensional PSF, which is 45 pixels wide. The bright source is small, nearly point-like, and much brighter than the other galaxy sources in the same scan, so it is a good match to the PSF.}
    \label{fig:source}
\end{figure}

The left-hand panel of Fig.~\ref{fig:source} shows the scan through the bright source again, along with an image of this source as an insert. The right-hand panel has an enlargement of the source profile as a red curve and, as a black curve, the average scan through a 2D PSF that was independently fit to our observations (as a 45$\times$45 pixel image) from known point sources outside the galaxy. The intensity profile of the bright source matches well the idealized PSF. 

\vspace{1cm}
\subsection{Power Spectra}
\label{sect:628_560_PS}
Power spectra were determined for each intensity scan by summing over the products of the intensity with sine and cosine functions. We experimented with a correction for the different intensities at the beginnings and ends of each scan, which are viewed by a Fourier transform as a sudden jump between neighboring pixels. In this experiment, we attached cosine functions to each end with values of zero at the far ends and values equal to the scan values where they attached. The wavelength of this cosine function was twice the scan length, making a half-wave modifier at each end and a new scan with twice the length of the original scan.  With this replacement, the intensity jump between the beginning and ending pixels contributes only to the Fourier transform at the lowest $k$ value. However, applications of this cosine-padding method to the scan with the bright source introduced too large a signal at small $k$, giving the PS of the PSF a slightly negative slope instead of zero slope without the cosine padding. Also, many of the scans have intensities close to sky values at the ends, so the end-to-end jumps viewed by the Fourier transform were no more severe than any other pixel-to-pixel jumps from noise. For these reasons we do not include cosine padding for the results presented here, nor padding an equal length to the side of the intensity strip with zeros, as in \cite{grisdale17}. The primary change to the intensity scans was to subtract the average value, giving zero average for each before the Fourier transform. The relative scan amplitudes were not changed, so the averages of PS presented below are weighted by the brighter scans. 

\begin{figure}
    \includegraphics[width=\columnwidth]{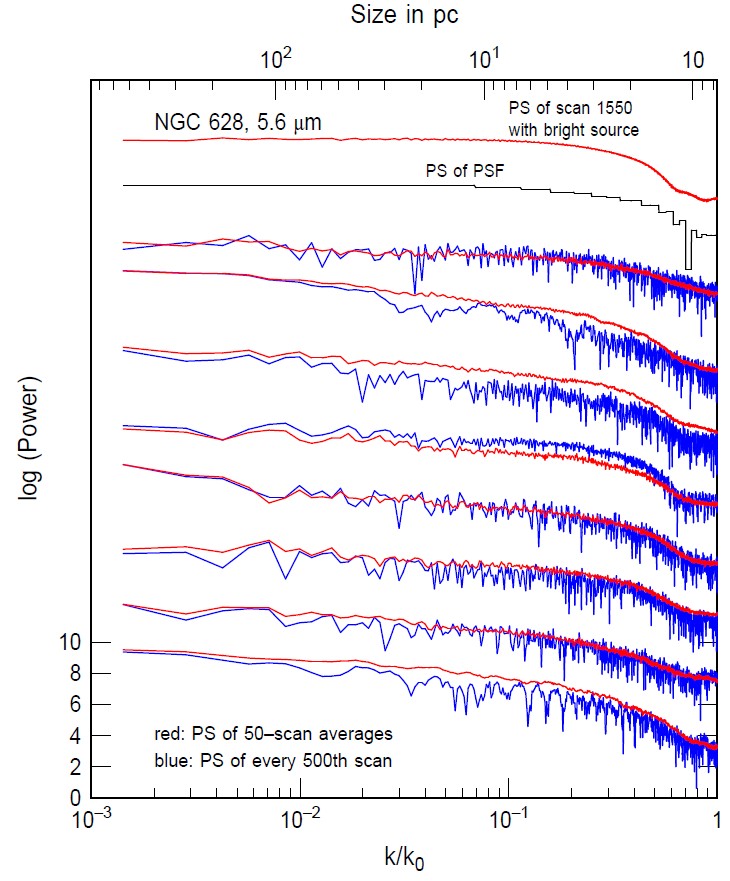}
    \caption{Blue: PS of the scans in Fig.~\ref{fig:intensity}, shifted vertically for better viewing.  Red: Average of 50 PS centered on the PS of the Fig.~\ref{fig:intensity} scans. The Fourier transform wavenumber is denoted by $k$, and the wavenumber corresponding to a wavelength of 2 pixels is denoted by $k_0$. The dip at high $k/k_0$ is the PSF from small bright sources in the image. The black PS at the top is the PS of the idealized PSF, which is 45 pixels wide and stretched to fit this scale for NGC 628. The red PS at the top is the PS of the single intensity scan through the small bright source indicated by a red circle in Fig.~\ref{fig:628image}.}
    \label{fig:ps628sample}
\end{figure}

The sine and cosine functions are evaluated by direct summations:
\begin{subequations}
\begin{align}
    S(k) &= \frac1{N} \sum_{m=1}^{N} I(m)\sin(2\pi km /N),\\
    C(k) &= \frac1{N} \sum_{m=1}^{N} I(m)\cos(2\pi km /N)
\end{align}
\end{subequations}
for wavenumber $k\in[1,N/2]$. The largest $k$ value corresponds to a wavelength consisting of 2 pixels, representing the two halves of the shortest complete sine wave. In what follows, we call this largest $k$ value $k_0$ and normalize the other $k$ values to it, plotting $k/k_0$. The power spectrum is
\begin{align}
    P(k) &= S^2(k)+C^2(k). 
\end{align}

Fig.~\ref{fig:ps628sample} shows in blue color the 8 PS that correspond to the 8 intensity scans in Fig.~\ref{fig:intensity}. Each PS is noisy so we average together the 50 PS surrounding these scans (from $\pm25$ strips) to make a single PS with higher signal-to-noise. This number of scans in the average was chosen to be large enough to reduce noise and allow a kink in the PS to be seen, and small enough to isolate distinct parts of the galaxy. The 50-PS averages are shown as red curves in Fig.~\ref{fig:ps628sample}. At the top of the figure are the PS of the scan with the bright source (red curve) and the PS of the idealized PSF (black curve).  There is PSF structure out to several arcsec that causes a dip in the PS at high $k$.  The lower x-axis is the relative wavenumber, $k/k_0$, and the upper x-axis shows the corresponding length scale in parsecs. A typical disk thickness of $\sim100-300$ pc would be in the center of the scan on this logarithmic scale. 

\begin{figure}
    \includegraphics[width=8.5cm]{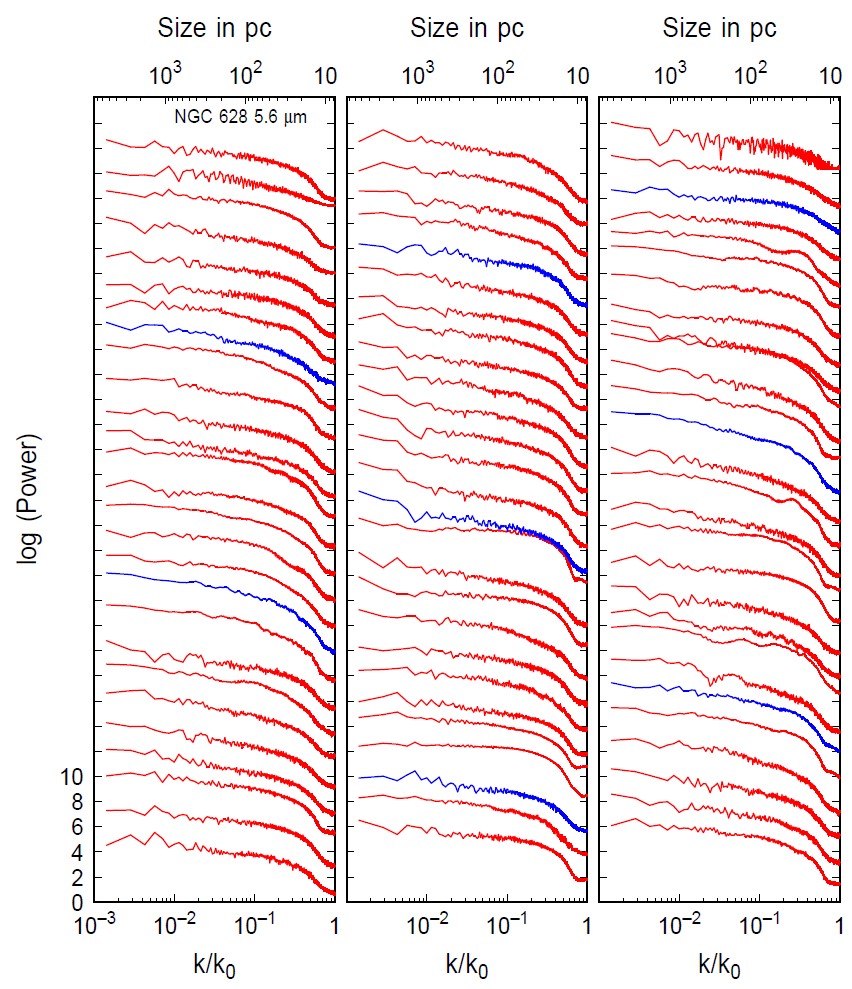}
    \caption{Averages of each 50 adjacent PS covering the image in Fig.~\ref{fig:628image}, shifted vertically for clarity. The 50-PS average corresponding to the southeast part of the image is at the bottom of the left-hand panel. Higher in the same panel, and then from bottom to top in the middle panel and bottom to top in the right-hand panel are the 50-PS averages in sequence from southeast to northwest in the image. The blue curves denote every 10th 50-PS average, which correspond to the positions of the blue lines in Fig.~\ref{fig:628image}.  }
    \label{fig:ps628_560_all}
\end{figure}

There is significant variation in the PS from scan to scan, reflecting the presence or lack of bright sources, including the disk itself for scans that go close to the galaxy center.  Fig.~\ref{fig:ps628_560_all} shows all of the 50-PS averages, from the southeast part of the galaxy in the lower left of the figure, to the northwest part in the upper right. The average PS are spaced vertically by a constant amount for clarity; every tenth one is blue.  They all have a dip at high $k/k_0$ from the PSF at $5.6\mu$m, but they vary in slope at low $k/k_0$. The bright source is in scan 1550, which is in the 31st average PS. This is the 4th curve up from the bottom in the central panel. Other PS averages with bright sources have the same shape as this one, namely a near-zero slope at low $k/k_0$ and a greater-than-average drop at high $k/k_0$ from the PSF. There is no evidence for a kink in any of the average PS at a wavelength corresponding to the suspected disk thickness of $\sim100-300$ pc. 

\begin{figure}
    \includegraphics[width=8.5cm]{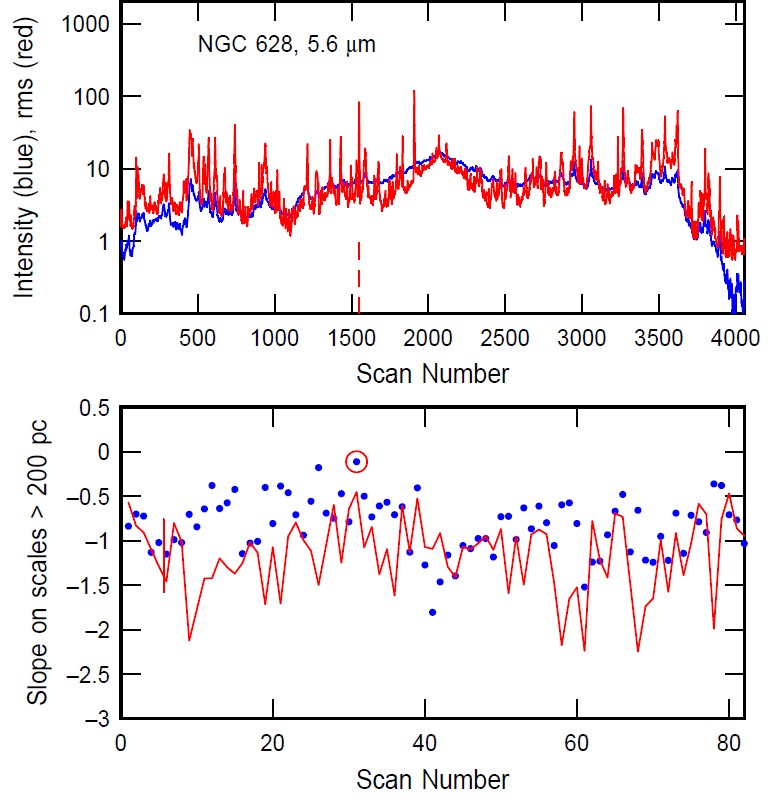}
    \caption{Bottom: Slopes of the low-$k$ parts of the 50-PS averages for scales larger than 200 pc, versus the scan number from 1 to 82 corresponding to scans from southeast to northwest in the galaxy image of Fig.~\ref{fig:628image}. Top: average intensity in units of $10^{-8}$ Jy px$^{-1}$ (blue) and rms (red) of each scan, in order from southeast to northwest. The dashed red line indicates the position of the bright source shown by a red circle in Fig.~\ref{fig:628image}. Corresponding to this is the point in the bottom panel with the red circle. The PS slope at low-$k$ is close to zero for the scan with the dominant point-like source. The PS slope systematically gets more negative for scans going through the galaxy center because of the disk exponential profile along the scan direction, which introduces significant power on large scales.}
    \label{fig:slope628_560}
\end{figure}

The bottom panel of Fig.~\ref{fig:slope628_560} shows the slopes of the 50-scan PS averages on scales larger than 200 pc  as a function of scan number. These slopes were determined from least-squares fits to the $\log P(k/k_0)$ versus $\log(k/k)0)$ relations in the plots. The average slope is $-0.83$ and the average error in the slope is $0.17$, as determined from the 90\% uncertainty limit in a student-t distribution. The error is smaller than the rms variance in the slope, which is $\pm0.31$. These values and analogous  values for the other wavelengths and galaxies are in Table 2. 

The top panel of Fig.~\ref{fig:slope628_560} shows the average intensities of all the scans, in blue, and the rms values around those averages, in red. The rms spikes when there is a bright source in the scan. The scan containing the brightest point-like source, scan 1550, is indicated by a vertical dashed line in the top panel and a red circle in the bottom panel. As was evident from previous figures, the slope of the PS of the bright source is $\sim0$ on large scales. Other spikes in the rms correspond to shallow slopes too. Fig.~\ref{fig:slope628_560} also shows a systematically steeper (negative) slope for scans that go through the inner part of the galaxy. These inner scans include the exponential disk profile which is a large-scale structure, and that adds power at low $k/k_0$, making the slope more negative.  More negative slopes here are directly evident in Fig.~\ref{fig:ps628_560_all} around the 40th PS-average, which is in the middle of the central panel. 

\begin{deluxetable}{cccccc}
\tabletypesize{\scriptsize}
\tablecaption{Power Spectrum Results \label{table:results}}
\tablewidth{0pt}
\tablehead{
\colhead{Galaxy}
&\colhead{Filter}
&\colhead{Scan}
&\colhead{50-PS Slope}
&\colhead{50-PS}
\\
NGC&&Direction&$>200$ pc&Slope rms
}
\startdata
628  & F560W  & Minor Axis & $-0.83\pm0.31$ & 0.17 \\
628  & F770W  & Minor Axis & $-0.86\pm0.26$ & 0.20 \\
628  & F1000W & Minor Axis & $-0.80\pm0.25$ & 0.19 \\
628  & F2100W & Minor Axis & $-0.63\pm0.34$ & 0.15 \\
5236 & F560W  & Minor Axis & $-1.17\pm0.59$ & 0.39 \\
5236 & F770W  & Minor Axis & $-1.11\pm0.51$ & 0.41 \\
4449 & F560W  & Minor Axis & $-0.95\pm0.43$ & 0.29 \\
4449 & F560W  & Major Axis & $-0.91\pm0.35$ & 0.23 \\
4449 & F770W  & Minor Axis & $-0.99\pm0.37$ & 0.37 \\
4449 & F770W  & Major Axis & $-0.92\pm0.33$ & 0.26 \\
5068 & F1000W & Minor Axis & $-0.80\pm0.36$ & 0.20 \\
5068 & F2100W & Minor Axis & $-0.58\pm0.37$ & 0.15 \\
\enddata
\end{deluxetable}

\vspace{1.5cm}
\section{Power Spectra of NGC 628 at $7.7$ microns, $10$ microns and $21$ microns}
\label{sect:628_770_1000_2100_PS}

Now we consider the $7.7\mu$m, $10\mu$m, and $21\mu$m images of NGC 628 taken with the F770W, F1000W, and F2100W filters as part of the FEAST and PHANGS-JWST programs. The pixel scales are still $0.08^{\prime\prime}$, but the nominal FWHM of the PSF is larger for longer wavelengths. For $7.7\mu$m, it is $0.269^{\prime\prime}$ or 3.36 pixels (12.8 pc),
for $10\mu$m is it $0.328^{\prime\prime}$ or 4.1 pixels (15.6 pc),
and for $21\mu$m, it is $0.674^{\prime\prime}$ or 8.425 pixels (32.1 pc, see Table 1).

\begin{figure}
    \includegraphics[width=\columnwidth]{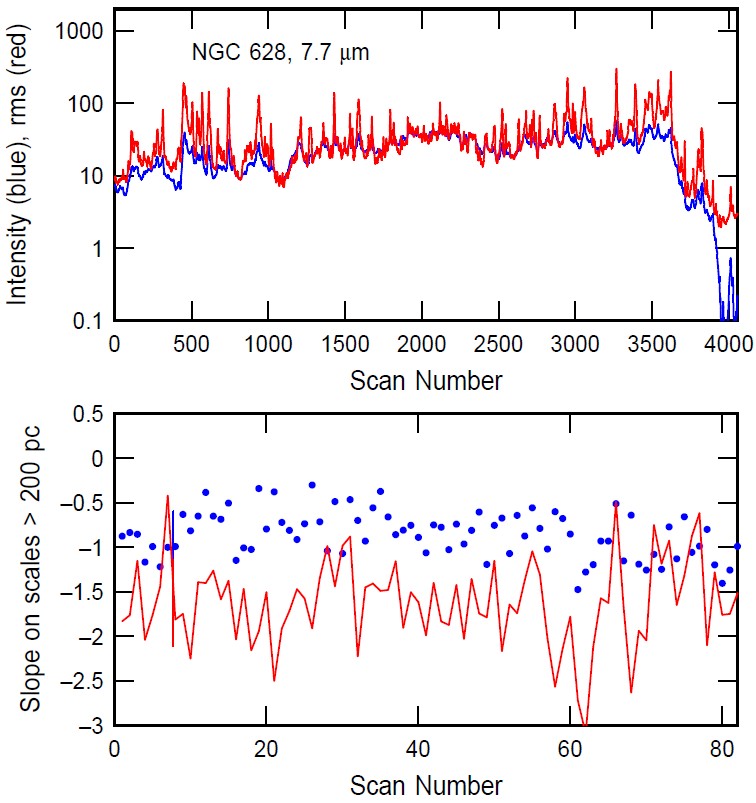}
    \caption{Same as Fig.~\ref{fig:slope628_560} but for the $7.7\mu$m image of NGC 628. }
    \label{fig:slope628_770}
\end{figure}

\begin{figure}
    \includegraphics[width=\columnwidth]{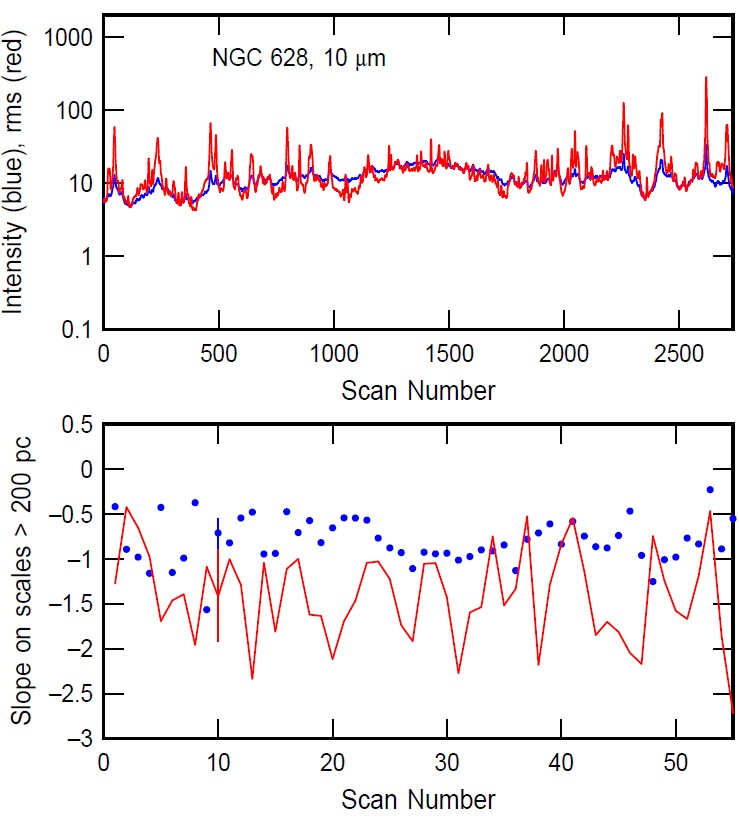}
    \caption{Same as Fig.~\ref{fig:slope628_560} but for the $10\mu$m image of NGC 628. }
    \label{fig:slope628_1000}
\end{figure}

Images for $7.7\mu$m and $10\mu$m with sample scan lines are in Figures \ref{fig:628_770_image} and \ref{fig:628_1000_image}, and 50-PS averages at these wavelengths are in Figures \ref{fig:ps628_770_all} and \ref{fig:628_1000_all}. The slopes on scales larger than 200 pc are in Table \ref{table:results}, and the mean intensities and rms values are in Figures \ref{fig:slope628_770} and \ref{fig:slope628_1000}, respectively. The intensity spikes are not as high at $7.7\mu$m as at $5.6\mu$m, and the large-scale PS slopes are not as close to zero for the highest intensity spikes. Neither is there a significant influence by the exponential disk in steepening the slope in the central region of the galaxy. On the other hand, there are a few strong spikes in the $10\mu$m image and one slope value close to zero at average-PS number 53 (lower panel on the right). There is also more evidence for the exponential disk in the 10$\mu$m PS going near the galaxy center. 

\begin{figure}
    \begin{center}
        \includegraphics[width=6cm]{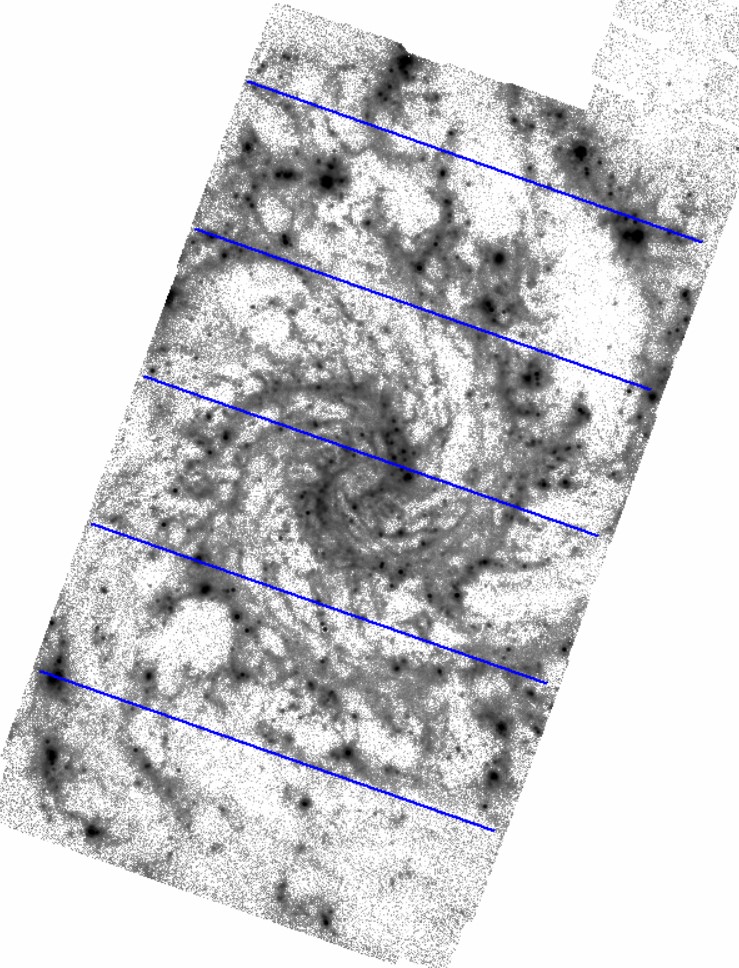}
    \end{center}
    \caption{NGC 628 at $21\mu$m with sample scan directions like those used to determine the PS. The separation between lines is 500 pixels, which is $40^{\prime\prime}$, or 1.91\,kpc.
    }
    \label{fig:628_21_image}
\end{figure}

\begin{figure}
    \includegraphics[width=\columnwidth]{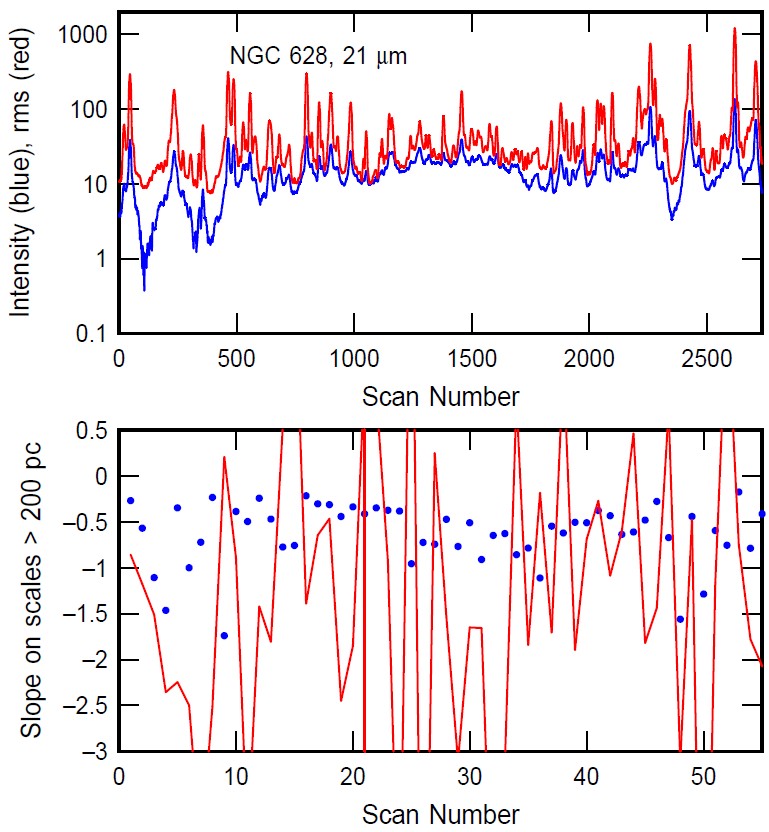}
    \caption{Same as figure \ref{fig:slope628_560} but for the $21\mu$m image of NGC 628.}
    \label{fig:slope628_2100}
\end{figure}

The $21\mu$m intensity is more spiked with bright sources than any of the others. Fig.~\ref{fig:628_21_image} shows the scan directions on the $21\mu$m image. The image looks significantly different than the $5.6\mu$m image in Fig.~\ref{fig:628image} because of the relatively brighter spiral arms, spurs and star-forming regions at $21\mu$m. These differences are also reflected in the 50-PS average slopes in Fig.~\ref{fig:slope628_2100}, which are generally closer to zero than at shorter wavelengths, and in the relatively high rms values compared to the average intensities in the top panel. All of the 50-PS averages are Fig.~\ref{fig:ps628_2100_all}. Corresponding to the intensity spikes, there are many flat PS with steep drops into the PSF depression. 

Comparing all of the 50-PS averages for NGC 628, there is a general trend toward flatter slopes on large scales for longer observed wavelengths, with slope averages (Table 2) varying as $-0.83\pm0.31$, $-0.86\pm0.26$, $-0.80\pm0.25$, and $-0.63\pm0.34$ for $5.6\mu$m, $7.7\mu$m, $10\mu$m and $21\mu$m, respectively. There is also an obvious extension of the PSF toward larger scales with longer wavelengths, even reaching the likely disk thickness close to $\sim100$ pc at $21\mu$m. Nevertheless, there is no indication of a systematic thickness kink in these PS averages: most of the structure seems to come from discrete sources. There are a few 50-PS averages that have slight PS kinks at $\sim200$ pc, but mostly the PS is curved in this range, rather than kinked. We comment on a possible origin for the $-0.6$ to $-0.9$ slopes at large scales in Section \ref{sect:discussion}. 

\section{Power Spectra of NGC 5236, NGC 4449, and NGC 5068}
\label{sect:5236_4449_5068}

A closer galaxy will have better spatial resolution than NGC 628. In the FEAST program, there is also the barred galaxy NGC 5236 at 4.66 Mpc \citep{tully13} and the dwarf galaxy NGC 4449 at 4.27 Mpc \citep{tully13} in $5.6\mu$m and $7.7\mu$m passbands. PHANGS-JWST \citep{lee23} has another relatively close dwarf galaxy, NGC 5068 at 5.2 Mpc \citep{anand21,leroy21} that has data at $10\mu$m and $21\mu$m. 

\begin{figure}
    \begin{center}
        \includegraphics[width=7cm]{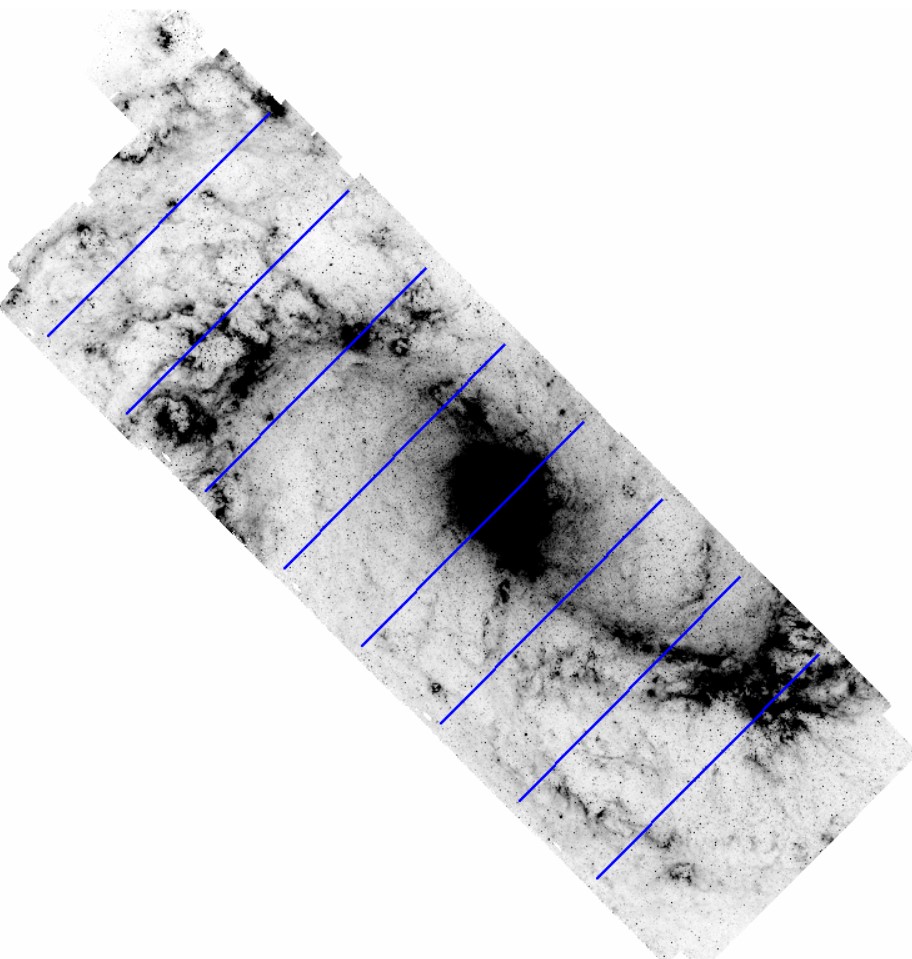}
    \end{center}
    \caption{NGC 5236 at $5.6\mu$m with sample scan directions like those used to determine the PS. The separation between lines is 500 pixels, which is $40^{\prime\prime}$, or 0.90\,kpc.}
    \label{fig:5236image}
\end{figure}

\begin{figure*}
    \hfil
    \includegraphics[width=8cm]{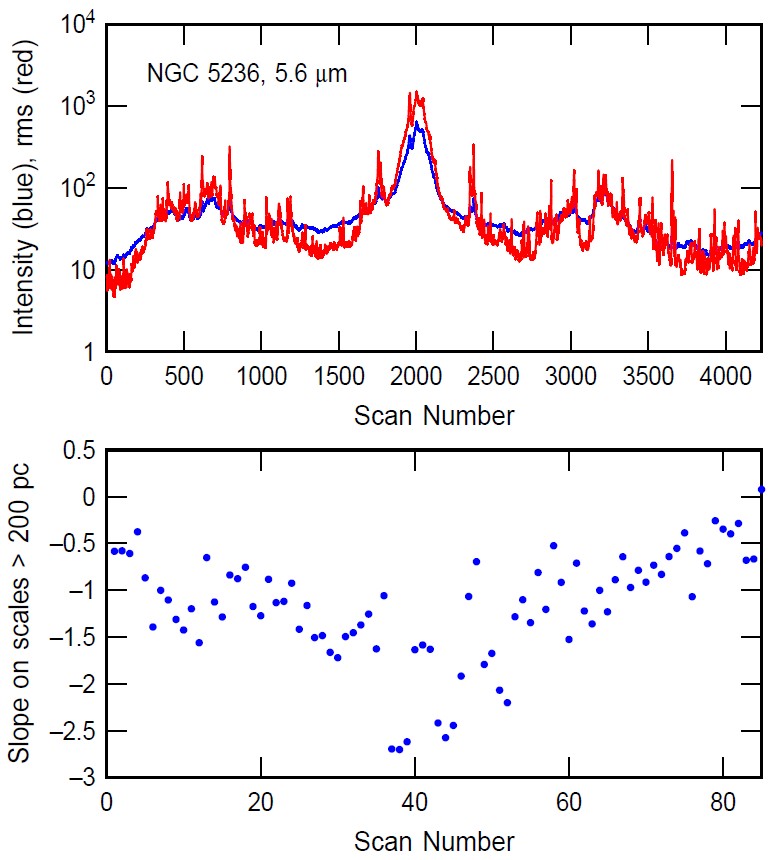}
    \hfil
    \includegraphics[width=8cm]{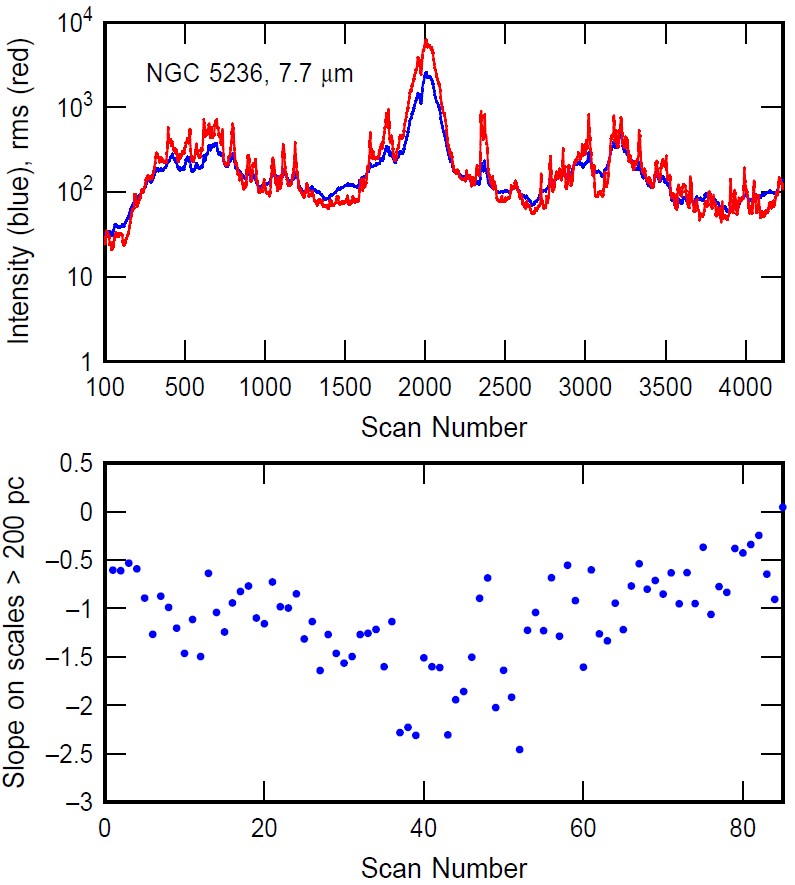}
    \hfil
    \caption{Low-$k$ slopes of the 50-PS averages (bottom) and average scan intensities (in $10^{-8}$ Jy px$^{-1}$, blue, top) and rms values (red, top) for NGC 5236 at $5.6\mu$ (left) and $7.7\mu$m (right). The slopes become more negative for scans through the inner regions of the galaxy because they contain the exponential disk profile which has power on large scales.}
    \label{fig:slope5236}
\end{figure*}

Scan directions through the $5.6\mu$m image of NGC 5236 are shown in Fig.~\ref{fig:5236image}, with scan numbers 500, 1000, 1500, etc, drawn as blue lines. The average intensities, rms values, and 50-PS average slopes are in Fig.~\ref{fig:slope5236} for $5.6\mu$m and $7.7\mu$m. This galaxy has a bright central region at both wavelengths where the PS slopes become steeper as a result of the exponential disk. Figs.~\ref{fig:ps5236_560_all} and \ref{fig:ps5236_770_all} show the 50-PS averages. There are few bright star forming regions that produce flat PS, and there is no systematic evidence for a kink in the PS indicative of disk thickness. 

\begin{figure*}
    \hfil
    \includegraphics[width=7cm]{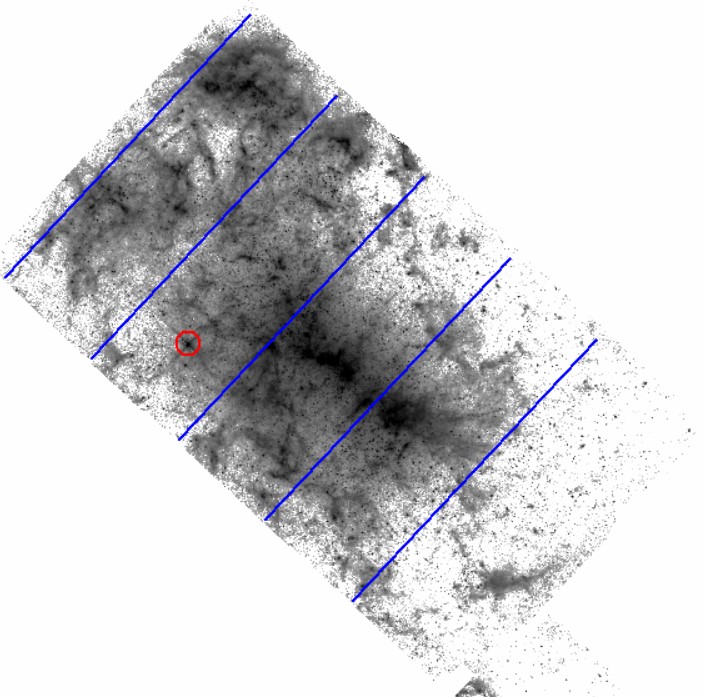}
    \hfil
    \includegraphics[width=7cm]{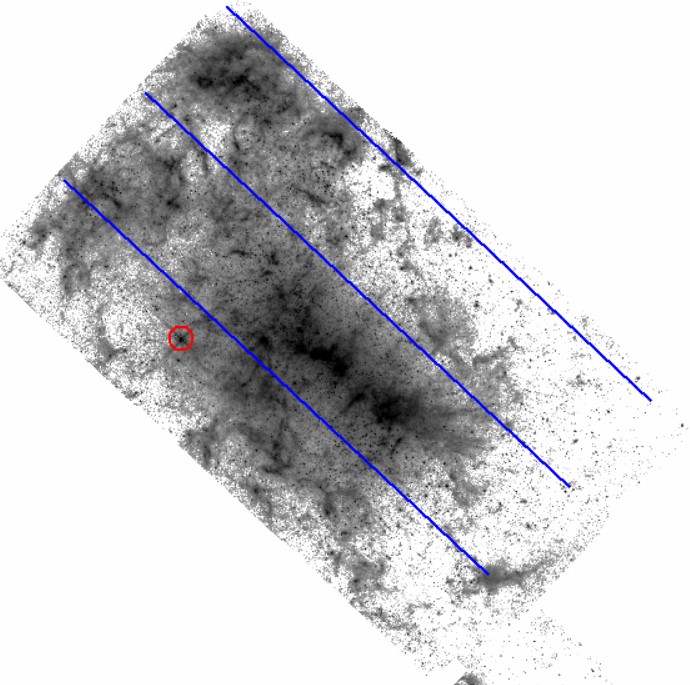}
    \hfil
    \caption{NGC 4449 at $5.6\mu$m with sample scan directions like those used to determine the PS. Sample minor axis scans are on the left and major axis scans are on the right. In both cases, the lines are separated by 500 pixels, which is $40^{\prime\prime}$ or 0.83\,kpc. The red circle denotes a bright source.}
    \label{fig:4449image}
\end{figure*}

\begin{figure*}
    \hfil
    \includegraphics[width=8cm]{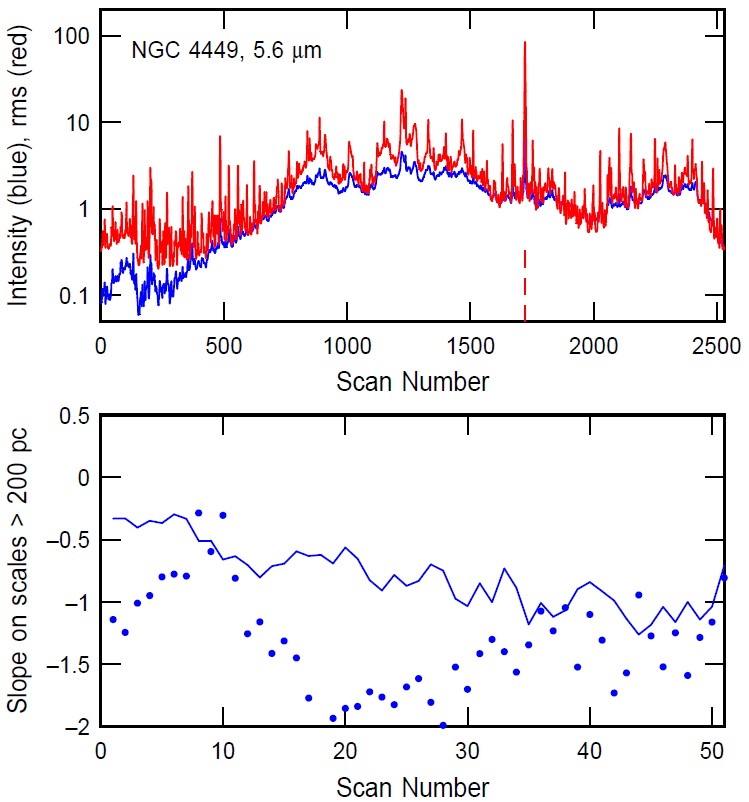}
    \hfil
    \includegraphics[width=8cm]{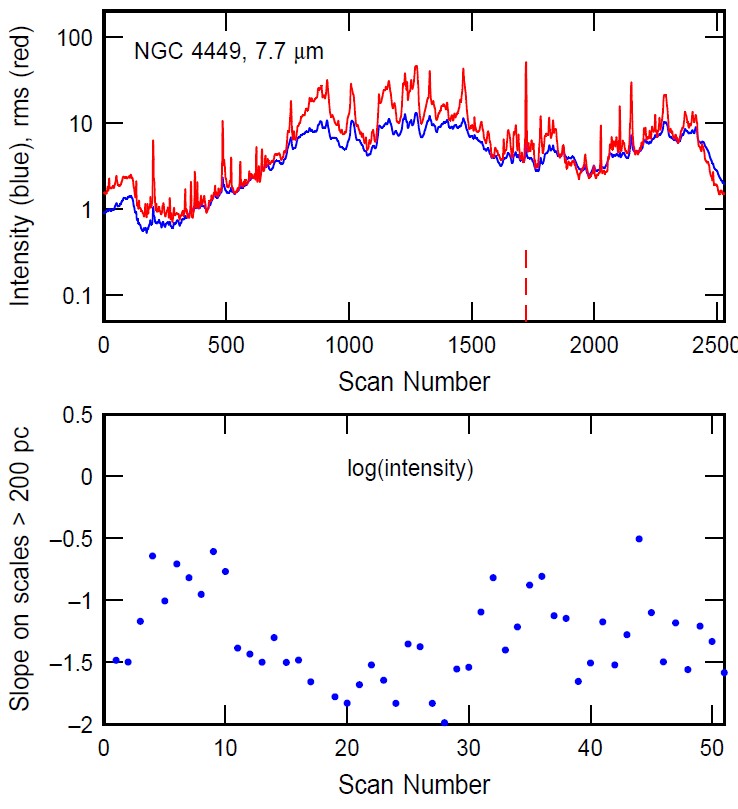}
    \hfil
    \caption{Low-$k$ slopes of the 50-PS averages (bottom) and average scan intensities (in $10^{-8}$ Jy px$^{-1}$, blue, top) and rms values (red, top) for minor axis scans though NGC 4449 at $5.6\mu$ (left) and $7.7\mu$m (right). }
    \label{fig:slope4449_minor}
\end{figure*}

\begin{figure*}
    \hfil
    \includegraphics[width=8cm]{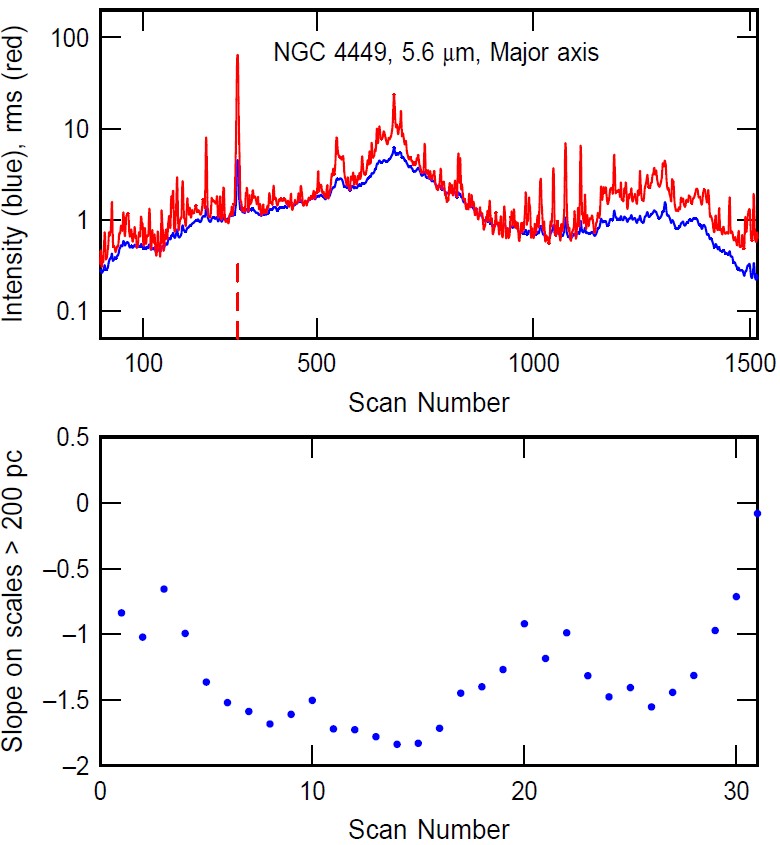}
    \hfil
    \includegraphics[width=8cm]{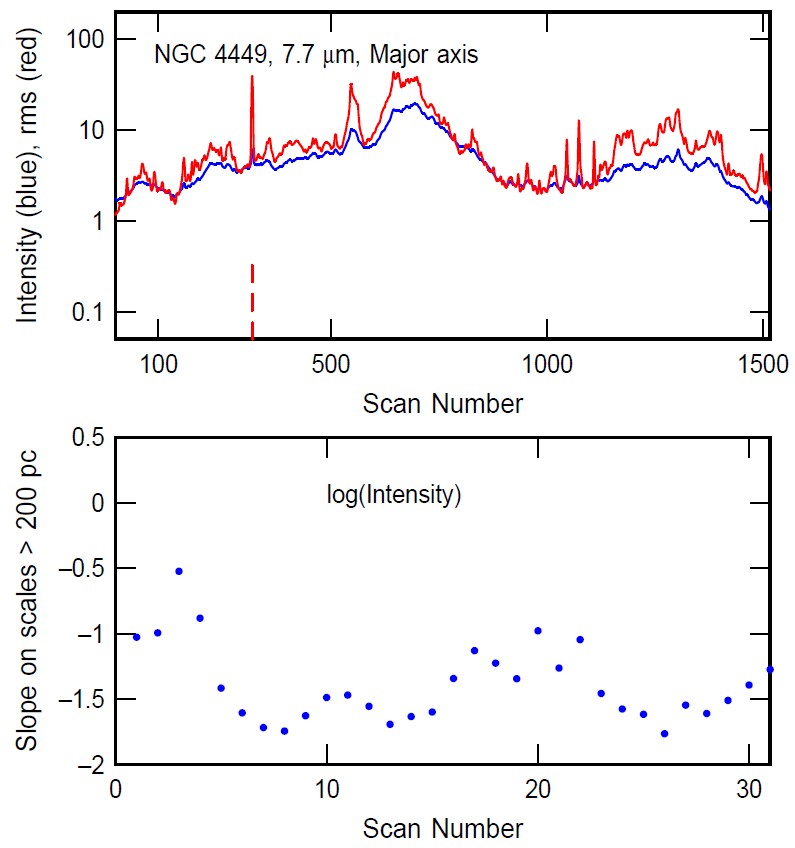}
    \hfil
    \caption{Low-$k$ slopes of the 50-PS averages (bottom) and average scan intensities (in $10^{-8}$ Jy px$^{-1}$, blue, top) and rms values (red, top) for major axis scans though NGC 4449 at $5.6\mu$ (left) and $7.7\mu$m (right).}
    \label{fig:slope4449_major}
\end{figure*}

Fig.~\ref{fig:4449image} shows NGC 4449 at $5.6\mu$m with minor and major axis scan directions to make PS with two different scan lengths (the blue lines are scan numbers 500, 1000, 1500, etc.). Figures \ref{fig:slope4449_minor} and \ref{fig:slope4449_major} show average intensities, rms values, and large-scale slopes for these two scan directions, respectively. A bright source indicated by the red circle in Fig.~\ref{fig:4449image} corresponds to the intensity peaks and slopes indicated by dashed lines and red circles in Figures \ref{fig:slope4449_minor} and \ref{fig:slope4449_major}. Again, the large scale slope of the PS is close to 0 for the scan with the bright point-like source, and the PS get steeper through the central regions. 

Figs.~\ref{fig:ps4449_560_all} to \ref{fig:ps4449_770_major_all} show the 50-PS averages in all 4 cases for NGC 4449. The bright source has the flat and smooth PS at the lower part of the right hand panel for the minor axis PS (Figs.~\ref{fig:ps4449_560_all}, \ref{fig:ps4449_770_all}), and the 7th PS up from the bottom of the left-hand panel for the major axis PS (Figs.~\ref{fig:ps4449_560_major_all}, \ref{fig:ps4449_770_major_all}). There is no systematic indication of a kink in the 50-PS averages at what might be the disk thickness.

\begin{figure}
    \includegraphics[width=8cm]{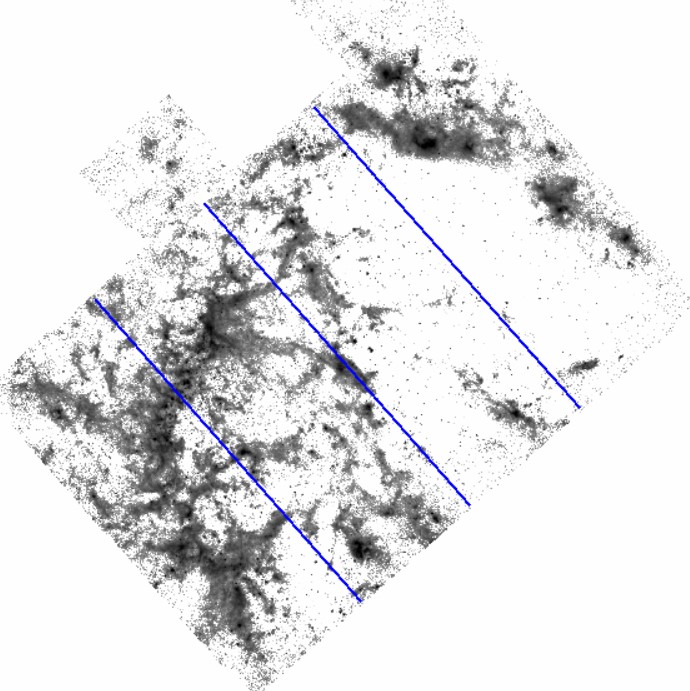}
    \caption{NGC 5068 at $10\mu$m with sample scan directions like those used to determine the PS. The separation between lines is 500 pixels, which is $40^{\prime\prime}$, or 1.00 kpc.}
    \label{fig:5068image}
\end{figure}

\begin{figure*}
    \hfil
    \includegraphics[width=8cm]{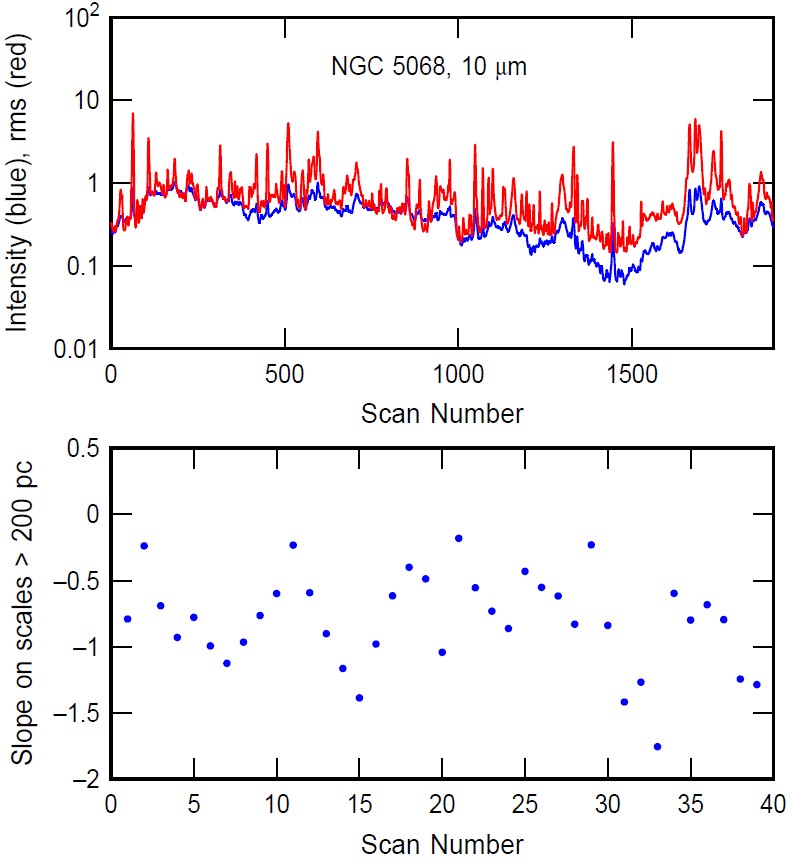}
    \hfil
    \includegraphics[width=8cm]{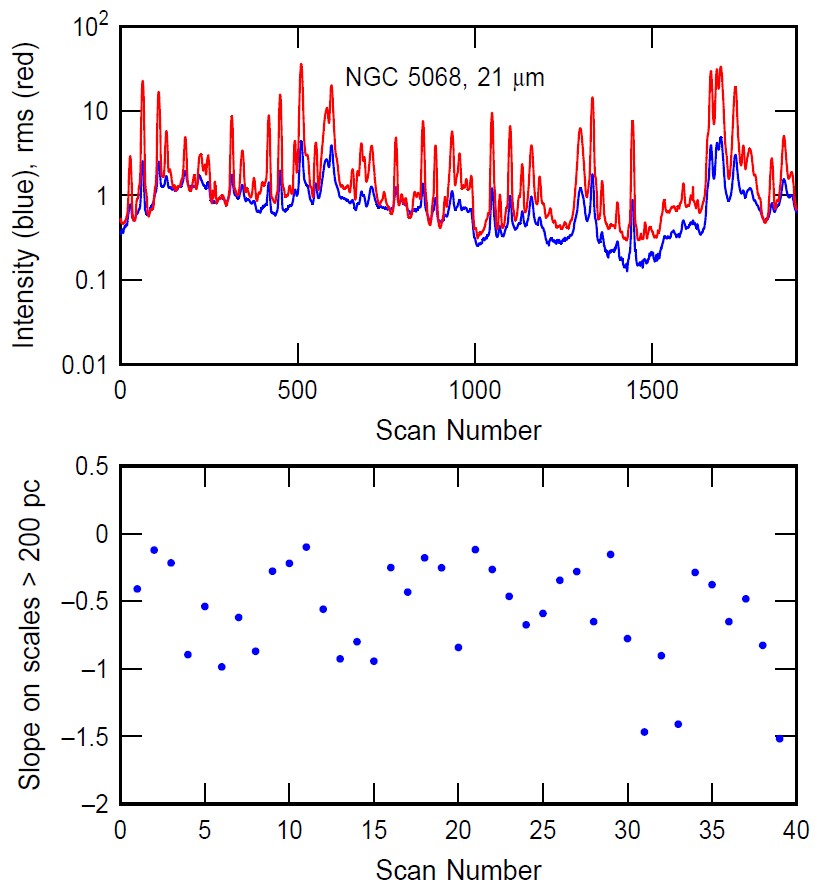}
    \hfil
    \caption{Low-$k$ slopes of the 50-PS averages (bottom) and average scan intensities (in $10^{-8}$ Jy px$^{-1}$, blue, top) and rms values (red, top) for  NGC 5068 at $10\mu$ (left) and $21\mu$m (right).}
    \label{fig:slope5068}
\end{figure*}

For NGC 5068, the pixel scale is $0.11^{\prime\prime}$. Fig.~\ref{fig:5068image} shows the scan directions for NGC 5068 on the $10\mu$m image and Fig.~\ref{fig:slope5068} shows the scan average intensities, rms values and large-scale slopes of the 50-PS averages. There are no particularly bright sources and no scans that clearly sample an exponential disk profile.  All of the 50-PS averages are in Figs.~\ref{fig:ps5068_1000_all} and
\ref{fig:ps5068_2100_all}. Some of these averages have the flat-slope signature of a scan with a primary point-like source, even though excessively bright intensities do not show up in Fig.~\ref{fig:slope5068}. As for the other galaxies studied here, there are no obvious systematic kinks in the PS that might indicate the disk thickness. 

\section{Power Spectra that avoid High Intensity Sources}
\label{sect:lowI}

PS variations from strong point-like sources prevent averaging over large regions of a galaxy because these averages will be contaminated by the few PS that are flat at low $k/k_0$.  To avoid this problem, we eliminated all intensity scans with peaks above certain thresholds, and then averaged the remaining PS together. We chose the thresholds to give around 30 good scans so the noise in the average PS is just low enough to reveal a possible kink. For NGC 628 images at $5.6\mu$m and $7.7\mu$m, these $\sim30$ scans are  among the 600 consecutive scans in each of six large regions. At $10\mu$m and $12\mu$m, the 30 select scans are out of 500 each in four large regions. The regions considered are shown by the colors in Figs.~\ref{fig:628_560_minor_image} and \ref{fig:628_1000_minor_image}; the lines are separated by 100 scans. We avoided the nuclear regions in each case. 

\begin{figure}
    \includegraphics[width=\columnwidth]{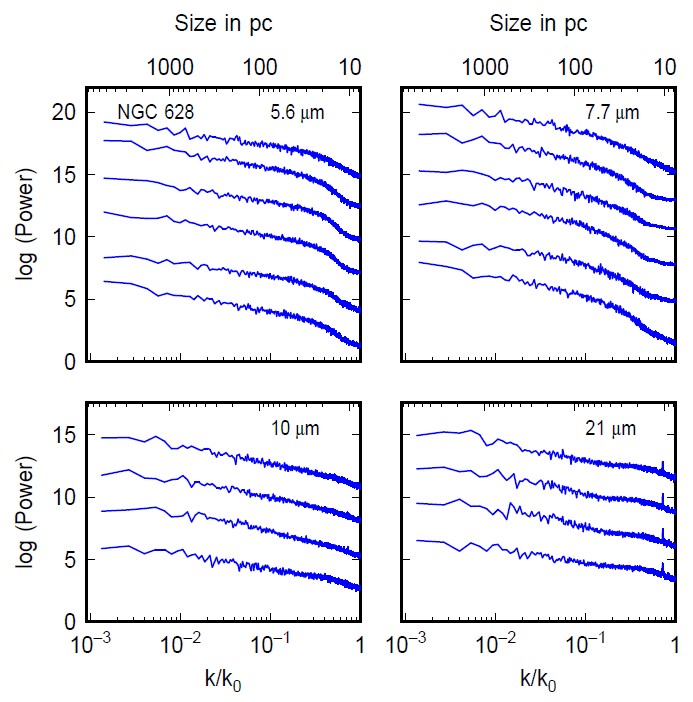}
    \caption{PS averages using only $\sim30$ intensity scans with the lowest peak intensities out of 600 scans, in each of 6 regions of NGC 628 for $5.6\mu$m and $7.7\mu$m and out of 500 scans for each of 4 regions for $10\mu$ and $21\mu$m. The regions are shown in Figs.~\ref{fig:628_560_minor_image} and \ref{fig:628_1000_minor_image}. The PS are plotted here from top to bottom in region order from north to south.}
    \label{fig:628_thresholds}
\end{figure}

Fig.~\ref{fig:628_thresholds} shows the 30-PS averages for the 6 regions in NGC 628 in the top panels, shifted by arbitrary amounts from bottom to top for regions in order from south to north ($5.6\mu$m is on the left and $7.7\mu$m is on the right). There is no evidence for a change in slope other than the dip from the PSF at high $k/k_0$. 

\begin{figure}
    \includegraphics[width=\columnwidth]{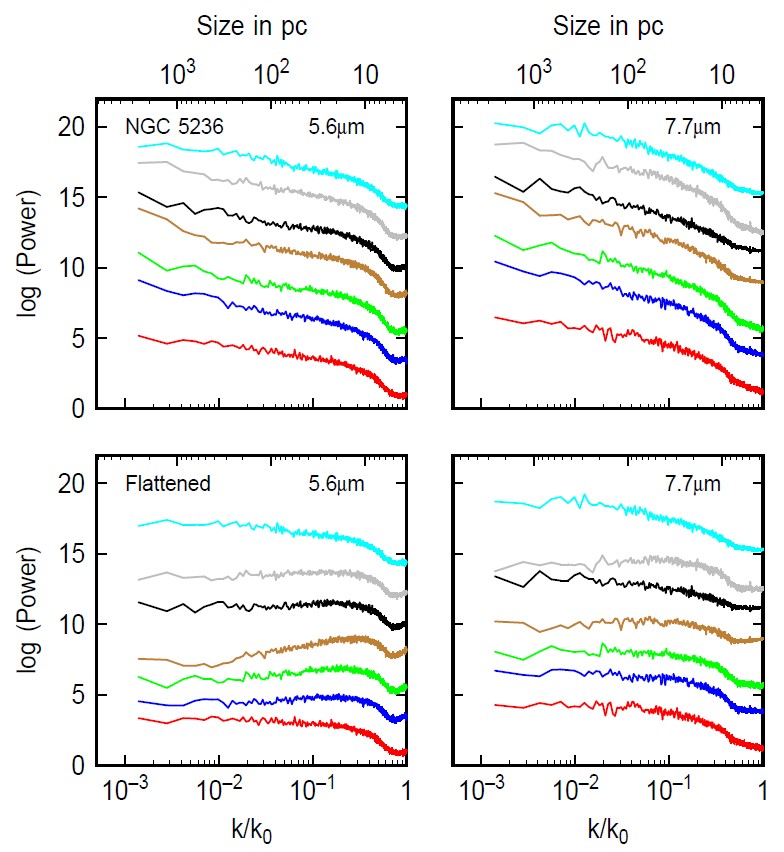}
    \caption{PS averages using $\sim30$ intensity scans with the lowest peak intensities out of 600 scans in each of 7 regions of NGC 5236 (not shown) for $5.6\mu$m and $7.7\mu$m. The PS are plotted from top to bottom in region order from northeast to southwest. The top panels show the PS and the bottom panels show the PS multiplied by a power of wavenumber that flattens the result to zero average slope on scales larger than 200 pc.}
    \label{fig:5236_thresholds}
\end{figure}

\begin{figure}
    \includegraphics[width=\columnwidth]{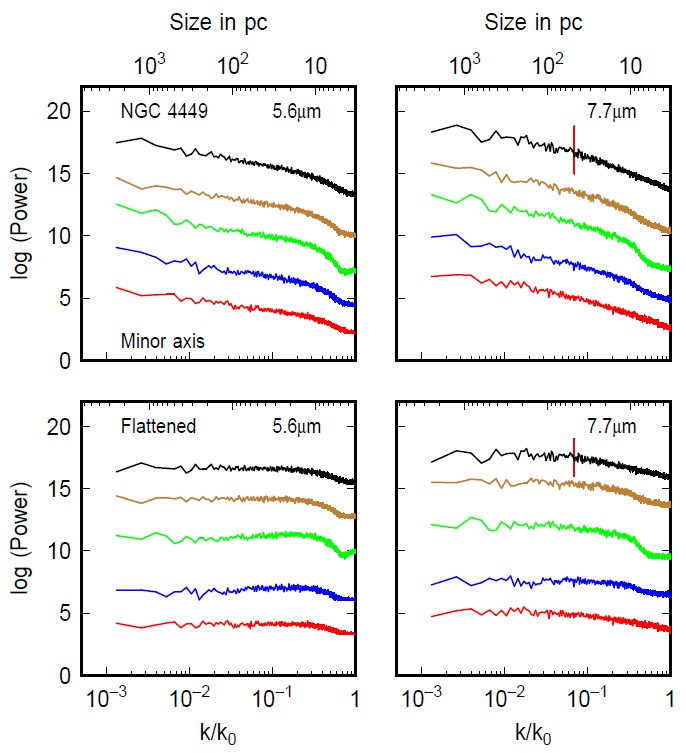}
    \caption{PS averages using $\sim30$ intensity scans with the lowest peak intensities out of 500 scans in each of 5 regions of NGC 4449 (not shown) for $5.6\mu$m and $7.7\mu$m. The PS are plotted from top to bottom in region order from northeast to southwest. The top panels show the PS and the bottom panels show the PS multiplied by a power of wavenumber that flattens the result to zero average slope on scales larger than 80 pc. There is a slight kink at 80 pc in the PS and fattened PS at $7.7\mu$m for the 500-scan region in the north.}
    \label{fig:4449_thresholds}
\end{figure}

\begin{figure}
    \includegraphics[width=\columnwidth]{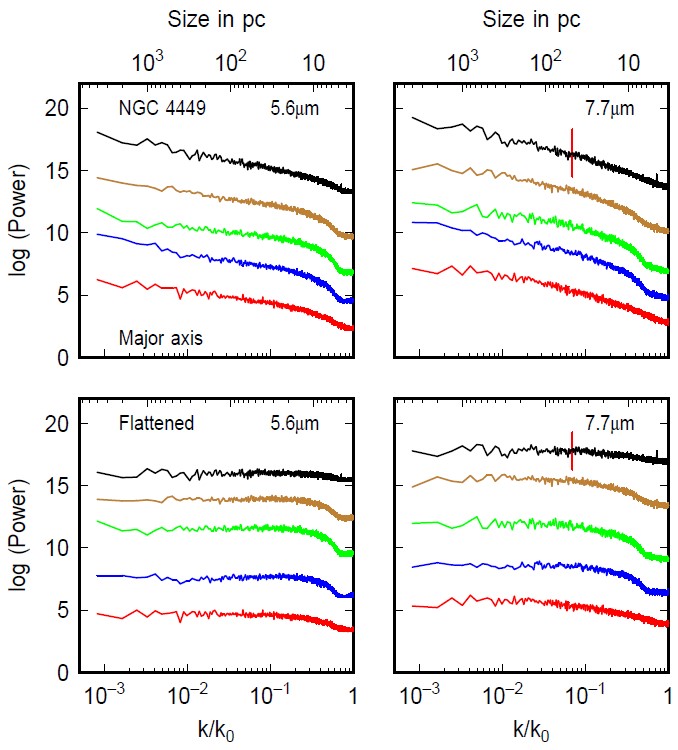}
    \caption{Same as Fig.~\ref{fig:4449_thresholds} using low-intensity scans but for major axis scan directions in NGC 4449. The order of the PS is from top to bottom in regions from northwest to southeast. There is a slight kink at 80 pc in the PS and fattened PS at $7.7\mu$m for the 300-scan region in the north.}
    \label{fig:4449_thresholds_major}
\end{figure}

\begin{figure}
    \includegraphics[width=\columnwidth]{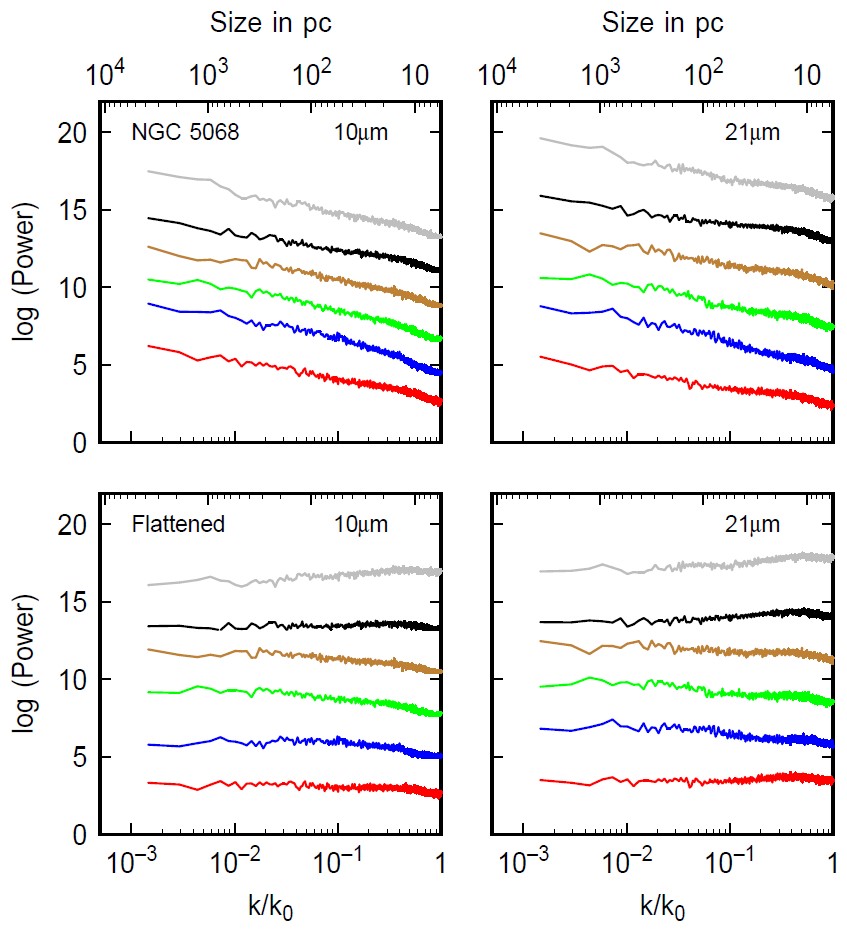}
    \caption{PS averages using $\sim30$ intensity scans with the lowest peak intensities out of 318 scans in each of 6 regions of NGC 5068 (not shown) for $10\mu$m and $21\mu$m. The PS are plotted from top to bottom in region order from northwest to southeast. The top panels show the PS and the bottom panels show the PS multiplied by a power of wavenumber that flattens the result to zero average slope on scales larger than 200 pc.}
    \label{fig:5068_thresholds}
\end{figure}

Similar intensity thresholding was done for the other three galaxies, with PS from $\sim30$ low-intensity scans averaged together in large regions, as shown in Figures \ref{fig:5236_thresholds}, \ref{fig:4449_thresholds}, \ref{fig:4449_thresholds_major}, and \ref{fig:5068_thresholds}.  For NGC 5236 there were 7 regions defined from north to south with 600 scans in each, for the NGC 4449 minor axis there were 5 regions of 500 scans each, for the NGC 4449 major axis there were 5 regions of 300 scans each, and for NGC 5068 there were 6 regions with 318 scans each. As for NGC 628, we eliminated all but around 30 scans from each of these regions and  average together the PS for these low-intensity scans. The top panels of these figures show the PS, and the bottom panels show the flattened PS, given by $k^{\alpha}P(k)$ where $-\alpha$ is the slope of the PS on scales larger than 200 pc for NGC 5236 and NGC 5068, and larger than 80 pc for NGC 4449. We choose 80 pc for NGC 4449 because there is a hint of a PS kink on that scale at $7.7\mu$. Flattening the PS at low $k/k_0$ makes a downward turn at high $k/k_0$ more visible. Aside from the $7.7\mu$m PS in the northern part of NGC 4449, the figures continue to show no kinks in the PS.

\section{Power Spectra from Azimuthal Profiles in NGC 628}
\label{sect:azimuthal}
As noted for Fig.~\ref{fig:slope628_560} and others in the preceding sections, the exponential radial disk profile steepens the low $k/k_0$ part of the PS by adding power on the largest scales. In this section we avoid that problem for NGC 628 by using azimuthal intensity scans instead of straight lines through the disk. Fig.~\ref{fig:628_1000_circles_image} shows circles at the positions of every 10th intensity scan, which is every 50th pixel (the scans are spaced apart by 5 pixels). The length of each azimuthal scan is much larger than the likely scale of the disk thickness, so the difference between a circular scan and a linear scan is negligible for our purposes. Also, NGC 628 is nearly face-on, so circles are a good approximation to azimuthal scans in the plane of the disk. 

\begin{figure}
    \includegraphics[width=\columnwidth]{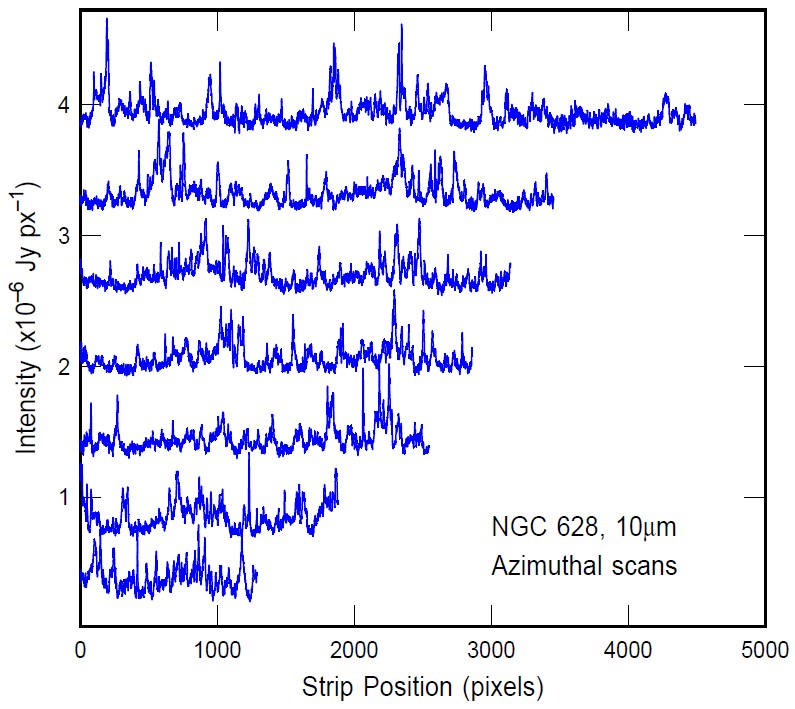}
    \caption{Intensity scans in the azimuthal direction for NGC 628 at $10\mu$m. These are a sample from the low-intensity subset of $\sim30$ azimuthal scans out of 145 total azimuthal scans that is used to avoid large variations and flattenings in the PS.}
    \label{fig:628_circles}
\end{figure}

\begin{figure}
    \includegraphics[width=\columnwidth]{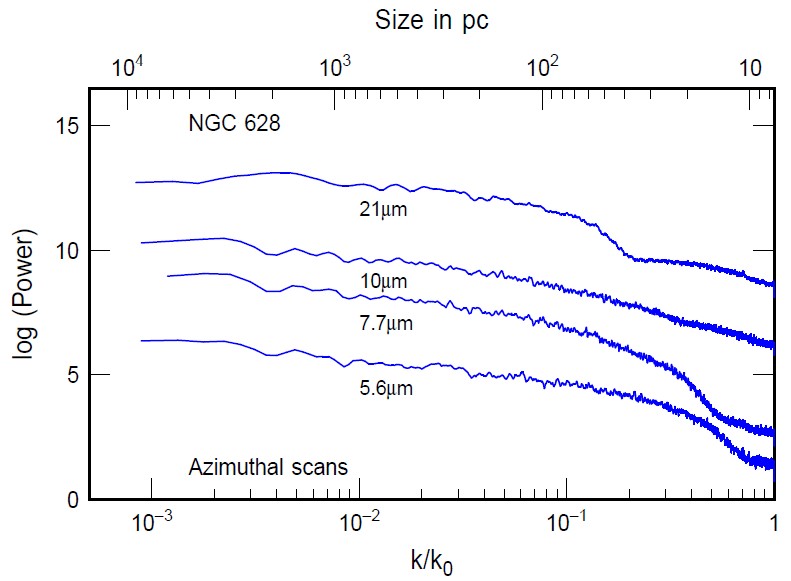}
    \caption{Average PS at 4 passbands from azimuthal intensity scans selected to have only low intensity peaks.}
    \label{fig:628_PS_circles}
\end{figure}

Fig.~\ref{fig:628_circles} shows select intensity scans from a low-intensity sub-sample. The length of the scan increases with radius. Average power spectra from the $\sim30$-scan sub-samples are in Fig.~\ref{fig:628_PS_circles}.  There is a gradual curvature to the PS which might indicate a transition from 2D to 3D turbulence, but no obvious kink that might indicate a disk thickness.

\section{Discussion}
\label{sect:discussion}
\subsection{Searching for a Thickness Kink in the PS}

The PS of face-on galaxies have the potential to show the line-of-sight thickness, where two-dimensional turbulence converts to three-dimensional turbulence, but there are severe constraints on the suitability of this method. The PS needs a reasonably large factor for the range of wavenumbers on either side of the thickness kink, which implies a spatial resolution at least a factor of $\sim10$ smaller than the galaxy thickness and a galaxy size at least a factor of $\sim10$ larger than the thickness. An additional constraint is that the observations must sample some type of emission that follows the correlated structure of turbulent gas.  These constraints were satisfied for HI and FIR emission from the LMC and M33, but they are difficult for other galaxies with the existing archive of data on the neutral component of the ISM. 

The present study uses JWST observations of dust emission at what is nominally sub-arcsec resolution from the full mirror diameter, which should be good enough to reach linear scales of several tens of parsecs for a large number of galaxies within 10 Mpc. However, the point spread function for the JWST image has a broad tail with a hexagonal shape from the mirror segments, and this PSF is $\sim5$ times larger than the nominal FWHM. The PSF is visible as a strong dip in the PS at high wavenumbers, since there is no signal received by the detector on smaller scales, only noise. For the furthest of our galaxies, NGC 628 at 9.84 Mpc, this dip extends as a broad depression in the PS to $\sim50$ pc ($5\times$FWHM) at $5.6\mu$m, and to $\sim64$ pc, $\sim78$ pc and $\sim160$ pc at $7.7\mu$m, $10\mu$m, and $21\mu$m wavelengths, respectively, putting the expected disk thickness slightly inside the PSF tail at the longest wavelengths. This problem with the PSF was also noted for the HI PS of galaxies in \cite{grisdale17}.

We experimented with removing the PSF by dividing the scan PS by the PS of the PSF. The results did not improve our sensitivity to a possible kink in the PSF wings because the PSF is a two-dimensional shape. Our one-dimensional intensity scans through the galaxy blend emission from neighboring scans in the PSF lobes, so the PS of the 2D PSF or the PS of a scan through the core of the PSF is not a good match to the contribution of the PSF to the scan's PS.

For the closer galaxies, NGC 5236, NGC 4449 and NGC 5068, we should be able to resolve and distinguish a kink in the PS if it occurs at the disk thickness, which may range from $\sim200$ pc for the spiral galaxy NGC 5236 to much larger values such as $\sim400$ pc for the dwarfs. This dwarf thickness corresponds to an intrinsic ratio of axis of $q_0=0.44\pm0.20$ \citep{johnson17} for a kpc-scale disk. Note that the separations between the blue lines in the images of these three galaxies are all about 1 kpc (see figure captions). Nevertheless, we do not see any obvious and systematic kinks in the 50-PS averages  (with the exception of a possible PS feature at 80 pc scale in the northern region of NGC 4449 observed at $7.7\mu$m; see Figs. \ref{fig:4449_thresholds} and \ref{fig:4449_thresholds_major}). The reason for this lack of an obvious kink could be that the PS varies too much from region to region. Extreme variations are partly the result of individual bright sources, which flatten the PS by dominating any fainter, possibly correlated structures from turbulence. Variations are also the result of the exponential disk profile, which steepens the PS because of its large scale in scans near the galaxy centers.

In Sections \ref{sect:lowI} and \ref{sect:azimuthal} we attempted to overcome these problems by limiting the scans to the $\sim5$\% with the lowest intensity peaks and by considering intensity scans in azimuth. There was still no obvious indicator of a disk thickness, although there was sometimes a smooth steepening of the slope toward higher $k/k_0$. 

\begin{figure*}
    \begin{center}
        \includegraphics[width=15cm]{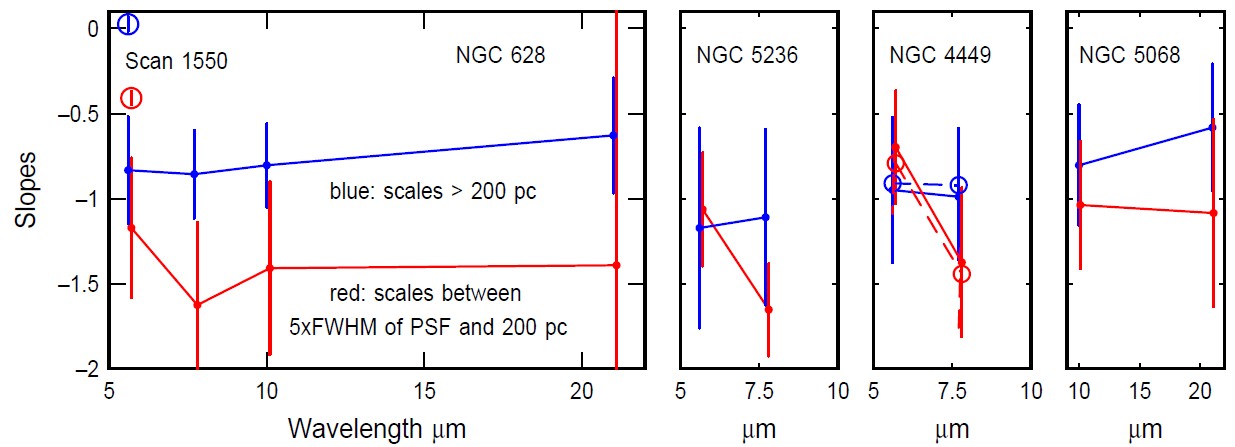}
    \end{center}
    \caption{50-PS average slopes for scales larger than 200 pc (blue) and for scales between 5 times the FWHM of the PSF (as listed in Table 1) and 200 pc (red) are shown versus the wavelength for all four galaxies. The rms range for each slope is shown by the error bars. Because 5 times the FWHM of the PSF at $21\mu$m in NGC 628 (which corresponds to $160.5$ pc) is close to 200 pc, there are very few $k/k_0$ values in this range and the rms is large, $\pm 2.1$, and mostly outside the figure. The two circles (with their tiny error bars) in the upper left corner are the slopes of the PS of scan 1550 for NGC 628 at $5.6\mu$m, which is the scan with the bright point-like source shown in Fig.~\ref{fig:source}. The red points are shifted to the right by $0.1\mu$m for clarity. }
    \label{fig:compare}
\end{figure*}

Fig.~\ref{fig:compare} examines more closely whether there is a slope change in the range from 100 pc to 300 pc that might be indicative of disk thickness. The blue curve is the average of all the 50-PS slopes on scales larger than 200 pc, and the red curve is the average of all the 50-PS slopes smaller than 200 pc, down to 5 times the FWHM of the PSF (Table 1), which is about the limit of the PSF tail. These slopes are plotted versus wavelength for the four galaxies. The large-scale slope is comparable to (or slightly more negative than) the small-scale slope at $5.6\mu$m, but at longer wavelengths, the PS slope is generally flatter on larger scales.  For NGC 628, the slope difference is about constant with wavelength beyond $5.6\mu$m, suggesting  that the increasing PSF width is not affecting the small-scale slope. The slope difference at these longer wavelengths is $\sim0.5$. The large-scale and small-scale slopes of the PS of scan 1550 in NGC 628, which contains the bright point-like source, are shown as circles in the upper left. As this scan best represents the effects of noise combined with the PSF, the slope difference is definitely not from disk thickness. Thus the other slope differences, which are about the same, are not likely to come from  disk thickness either. 

Aside from extreme variations in the PS with position, another explanation for the lack of an obvious kink is that most of the emission sources at the wavelengths considered here are closer to the midplane than expected. Previous PS with kinks tended to be based on diffuse gas that is traced by HI and cooler dust, and diffuse gas can be fairly thick. For example, the average hydrostatic equilibrium scale height for HI in 20 galaxies from The HI Nearby Galaxy Survey (THINGS) equals 330 pc \citep{bagetakos11}.  In contrast, the star-forming regions seen in our passbands are probably connected with dense molecular clouds, which are typically close to the midplane.  

The molecular cloud layer in the Milky Way has a FWHM thickness of $\sim100$ pc in the main disk \citep{heyer15}, which corresponds to a scale height, $\sigma_{\rm z}$, of 42 pc for a Gaussian profile. This height is in agreement with simulations of a Milky Way type galaxy by \cite{jeffreson22}, who got a molecular cloud scale height of 50 pc. In other measurements, \cite{scoville93} determined a FWHM of 220 pc ($\sigma_{\rm z}=84$ pc) in the edge-on galaxy NGC 891 using CO interferometry. This corresponds well with the half-width at half-maximum (HWHM) of 105 pc ($\sigma_{\rm z}=88$ pc) found by \cite{elmegreen20} for 133 star-forming cores in NGC 891 observed at $8\mu$m with the Spitzer Space Telescope InfraRed Array Camera. \cite{patra18} assumed hydrostatic equilibrium to derive the molecular HWHM in the highly inclined spiral galaxy NGC 7331, obtaining 50 pc - 80 pc at 6 kpc radius ($\sigma_{\rm z}=42$ pc - 67 pc).  For NGC 6946, \cite{patra21} measured a thin molecular HWHM of $\sim50$ pc ($\sigma_{\rm z}=42$ pc) at 4 kpc radius, and found a second molecular disk that is thicker by a factor of 2 and coincides with the HI disk. \cite{pety13} previously found a similar two-component molecular disk in M51, with half the molecular emission at least 10 times thicker than the dense molecular gas. The molecular thicknesses in Ultra Luminous Infrared Galaxies seem to be larger than those in normal galaxies, perhaps because of additional forcing from galaxy interactions; \cite{wilson19} found an average hydrostatic equilibrium thickness for CO of $\sim180$ pc in these galaxies. 

Evidently, the molecular thicknesses in normal spiral galaxies correspond to inverse wavenumbers that can blend with the tail of the PSF, masking a slight change in PS slope.  Molecular thicknesses in dwarfs are not as well known because CO emission is usually weak. 

\subsection{The Large-Scale Slope}
\subsubsection{A Turbulence Interpretation}
The slope of the large-scale part of the PS contains information about the emission structure. These slopes range from approximately $-0.6$ at $21\mu$m to  $-1.2$ at $5.6\mu$m for one-dimensional intensity strips (Table \ref{table:results}). 

Simulations of a turbulent galaxy disk by \cite{bournaud10} have a 2D power spectrum for surface density with a slope of $-1.9$ at small $k$ and $-3.1$ at large $k$. The one-dimensional analog for a strip of intensity in our case would be shallower by 1, namely, $-0.9$ and $-2.1$, respectively. This value of $-0.9$ at low $k/k_0$ resembles the slopes in Table \ref{table:sample}.  

Simulations of galaxies without feedback in \cite{fensch23} have a much shallower 2D PS, with a slope of $-0.7$ in the three-dimensional medium, i.e., at high $k$.  Their gas seems highly clumped, however, as feedback is needed to break apart the clumps formed by gravity \citep{bournaud10}. \cite{grisdale17} simulated PS for galaxy disks and found that a uniform background of HI increases the PS at low $k$, steepening the slope.  Our near-infrared emission does not have a significant background so this should not apply here. 

In the ideal case, the 2D PS of intensity integrated over a sufficiently deep sub-sonically turbulent layer will have the Kolmogorov slope of $-8/3$ on large scales and $-11/3$ on small scales \citep[][Fig.~3]{kolmogorov1941,lazarian00}.  Supersonic turbulence could have a slightly steeper slope \citep{burgers48}. Observations of 2D PS on large scales in many galaxies are roughly consistent with this, as summarized in Section [\ref{sec-intro}.   For 1D intensity strips like we have here, the Kolmogorov slopes become $-5/3$ and $-8/3$. This is approximately what was found for the 1D PS of HI emission in the LMC \citep{elmegreen01}.  In the infrared studied here, the low-$k$ PS are shallower than any of these, $\sim-0.9$ in Table \ref{table:sample}. 

If density is a passive tracer of turbulence, as proposed by \cite{goldman00} for the SMC, then diffusion can affect the slope of the PS. In the Batchelor regime on scales smaller than the dissipation length and larger than the diffusion length \citep{bedrossian21}, the power spectrum is $P(k)\propto k^{-1}$, closer to our value, in contrast to the inertial range on scales larger than the dissipation length, where $P(k)\propto k^{-5/3}$ for the Kolmogorov regime. Most likely the observed ISM is much larger than the dissipation length, so this shallow solution does not apply.  On scales larger than the dissipation length but shorter than the diffusion length, the power spectrum drops off exponentially toward higher $k$ because small scales diffuse before they convect \citep{tennekes72}. This rapid drop-off is not observed here either.


\subsubsection{An Interpretation based on Hierarchical Size and Luminosity Distribution Functions}

A different way to think about the PS is as a size-dependent distribution function of hierarchical emission sources. 
Many interstellar structures originate as point-like pressure sources, such as supernovae, HII regions, stellar winds, extragalactic impacts, etc., and the resulting morphologies consist of expanding bubbles and shell-like pieces around them. There is also a significant influence by self-gravity and galactic shear, which makes structures in the form of spiral arms and filaments. In this case, there is a different interpretation of galaxy-wide PS that is independent of conventional turbulence.  This different interpretation is not entirely without turbulence, as that would be present as well, but may be viewed as a superposition of definite structures inside a turbulent medium, with more turbulence inside the shells and spiral arms, as distinct from the shells and arms themselves.

With this source-based structure in mind, we consider now that $P(k)$ is a distribution function in the interval $k$ to $k+dk$ that arises from the combined (summed) luminosity $L(k)$ from all regions larger than $1/k$. This sum is because the emission coming from a particular region of size $1/k$ is the emission from that region plus the emission from the slightly larger region it is part of, plus the emission from the even larger region that both are from, etc., up to regions of (effectively) infinite size.  This consideration is analogous to the turbulent power spectrum, $P_{\rm turb}(k)$, which must be integrated over all wavenumbers $k$ to give the energy spectrum, $v^2\sim E(k)=P(k)\diff^3k\propto k^{-2/3}$ in Kolmogorov turbulence, where $P(k)\propto k^{-11/3}$ in three dimensions. The analogy arises because the total motion of a specific local region is the superposition of all the motions of the larger regions that enclose it. 

To apply this concept in the present case, i.e., viewing hierarchical interstellar structures in terms of the number distribution of objects and the luminosities of each, we convert the power spectrum as a function of wavenumber, $P(k)\diff k$, to an emission distribution as a function of size, $P_{\rm em}(R)\diff R$. We consider that $P(k)=P_{\rm em}(R) \diff R/\diff k\propto P_{\rm em}(R)k^{-2}$ for $R=1/k$.

We then write
\begin{align}
    P_{\mathrm{em}}(R)=\int_R^\infty L(R') \,n(R')\,\diff R'.
    \label{eq:pr}
\end{align}
where $L(R)$ is the luminosity of regions specifically of size $R$, i.e., subtracting their underlying backgrounds from the larger regions they are part of, and $n(R)\diff R$ is the size distribution of these regions. 

As an example of $L(R)$ and $n(R)$, we use the observation of superbubbles in four galaxies by \citet{oey97}.  They find a relationship between the luminosity function, $\Phi(L)\diff L\propto L^{-\beta}\diff L$, and the size distribution, $n(R)\diff R\propto R^{1-2\beta}$, where the same $\beta$ applies to each, almost  independently of evolutionary effects.  Converting these functions to a luminosity-size relation, $L(R)$, through the equivalence $\Phi(L)\diff L=n(R)\diff R$, we get
\begin{align}
    L^{-\beta}\diff L=R^{1-2\beta}\diff R,
\end{align}
from which it follows that $L\propto R^2$. For M33, they measure $\beta\sim1.6$ for scales up to a kpc. 

Using this notation for equation~\eqref{eq:pr}, we might write
\begin{align}
    P_{\rm em}(R)\propto \int_R^\infty  R'^2 R'^{1-2\beta}\diff R',
\end{align}
except that we derived our PS from one-dimensional scans, which makes $L_{\rm 1D}(R)\propto R$ instead of the $L\propto R^2$ measured by \citet{oey97} for whole regions. That is, if we picture the PS of a 1D intensity trace, an emission region contributes to that single scan only proportional to its length along the scan, unlike the total luminosity of the region which comes from its area and many adjacent 1D scans. Thus we have in our case for 1D intensity scans,
\begin{align}
    P_{\rm em}(R) \propto \int_R^\infty  R' R'^{1-2\beta}\diff R' \propto R^{3-2\beta}\sim R^{-0.2}
\end{align}
where the evaluation is again for $\beta=1.6$ from \citet{oey97}.

Converting this result to a function of $k$, we get
\begin{align}
    P(k) =P(R)\frac{\diff R}{\diff k} \propto R^{3-2\beta}k^{-2} = k^{2\beta-5}\sim k^{-1.8},
\end{align}
and converting further to equal intervals of $\log k$ instead of $k$, we get
\begin{align}
    P(k)\diff\log k\propto k^{2\beta-4}\diff\log k\sim k^{-0.8}\diff\log k,
\end{align}
which is the power spectrum observed at large $k$ in this paper (i.e., $\alpha\sim0.8$ to 1 from Table 2).

We note that if we were to take a two-dimensional PS, then we would use $L\propto R^2$ in the integral over size, giving $P_{\rm em}\propto R^{4-2\beta}$, which converts to $P(k)\diff k\propto k^{-2.8}\diff k$ or a slope on a log-log plot of $-1.8$. This is approximately the slope derived by others for 2D PS of galaxies, as summarized in the Introduction.

\section{Conclusions}
\label{sect:conclusions}
Power spectra of three galaxies in the FEAST survey and one galaxy in the PHANGS survey, using infrared wavelengths from $5.6\mu$m to $21\mu$m observed by JWST (Table \ref{table:sample}), show no clear evidence for disk thickness in the form of an inflection point from the transition between two-dimensional turbulence on scales larger than the thickness and three-dimensional turbulence on scales smaller than the thickness. This lack of a clear inflection is partly the result of large variations in the PS from position to position, reflecting discrete bright sources which tend to flatten the large-scale PS, and the exponential profiles of the galaxy disks, which tend to steepen the large-scale PS. In addition, the star-forming layers in these galaxies could be composed of dense gas that is several times thinner than the diffuse lower-density gas where PS kinks were previously found in different galaxies. This suggests that PS kinks for our galaxies could be blended with the extended tails of the JWST PSF. 

The PS slopes on large scale could be evidence for a source luminosity dependent on size, $L(R)$, and a source distribution function dependent on size, $n(R)\diff R$.  For source properties analogous to superbubbles observed in galaxies up to kpc scales \citep{oey97}, and for a hierarchical distribution of these sources, the observed PS slopes are reproduced.

\vspace{0.5cm}
{\it Acknowledgements:} B.G.E. would like to thank Alex Lazarian and the reviewers, Alessandro Romeo and an anonymous referee, for comments. This work is based in part on observations made with the NASA/ESA/CSA James Webb Space Telescope, which is operated by the Association of Universities for Research in Astronomy, Inc., under NASA contract NAS 5-03127. These observations are associated with program \#1783. Support for program \#1783 was provided by NASA through a grant from the Space Telescope Science Institute, which is operated by the Association of Universities for Research in Astronomy, Inc., under NASA contract NAS 5-03127. The authors acknowledge the team of the ``JWST-HST-VLT/MUSE-ALMA Treasury of Star Formation in Nearby Galaxies,''  led by co-PIs J. C. Lee, K.L. Larson, A.K. Leroy, K. Sandstrom, E. Schinnerer, and D.A. Thilker, for developing the JWST observing program \# 2107 with a zero-exclusive-access period. Archival data presented in this paper were obtained from the Mikulski Archive for Space Telescopes (MAST) at the Space Telescope Science Institute. K.G. is supported by the Australian Research Council through the Discovery Early Career Researcher Award (DECRA) Fellowship (project number DE220100766) funded by the Australian Government and by the Australian Research Council Centre of Excellence for All Sky Astrophysics in 3 Dimensions (ASTRO 3D), through project number CE170100013. A.A. and A.P. acknowledge support from the Swedish National Space Agency (SNSA) through grant 2021-00108. A.A. acknowledges support by the Swedish research council Vetenskapsr{\aa}det (2021-05559). M.M. acknowledges financial support through grant PRIN-MIUR 2020SKSTHZ. A.D.C. acknowledges the support from a Royal Society University Research Fellowship (URF/R1/191609).

{\it Facility:} JWST (NIRCam, MIRI). Software: JWST Calibration Pipeline \citep{bushouse22, greenfield16}; SAOImage DS9 \citep{joye03}; FORTRAN, and VossPlot \citep{voss95}.

\bibliography{refs}
\bibliographystyle{mnras}

\begin{appendix}
\section{Additional Figures}

\begin{figure}[H]
    \begin{center}
        \includegraphics[width=6.5cm]{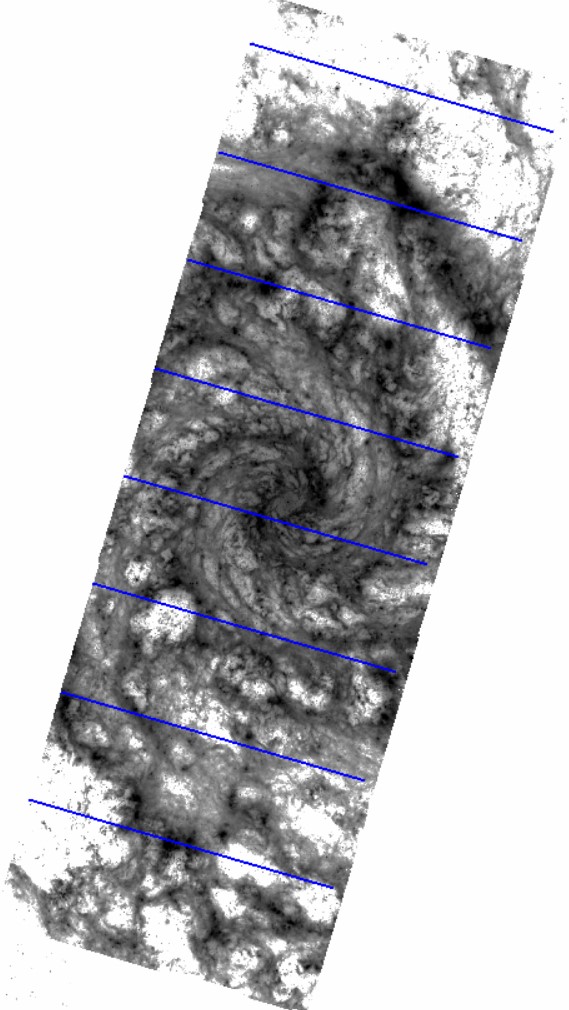}
    \end{center}
    \caption{NGC 628 at $7.7\mu$m with sample scan directions   like those used to determine the PS, as in Fig.~\ref{fig:628image}.
    }
    \label{fig:628_770_image}
\end{figure}

\begin{figure}[H]
    \includegraphics[width=8.5cm]{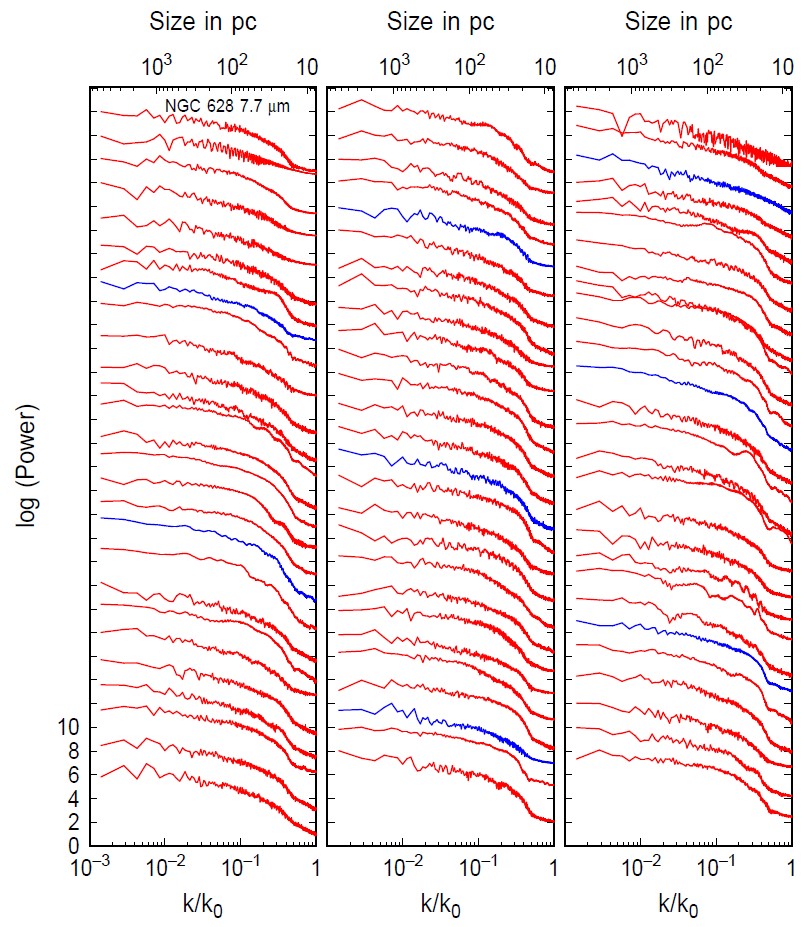}
    \caption{The same as Fig.~\ref{fig:ps628_560_all} but at $7.7\mu$m. }
    \label{fig:ps628_770_all}
\end{figure}

\begin{figure}[H]
    \begin{center}
    \includegraphics[width=6.5cm]{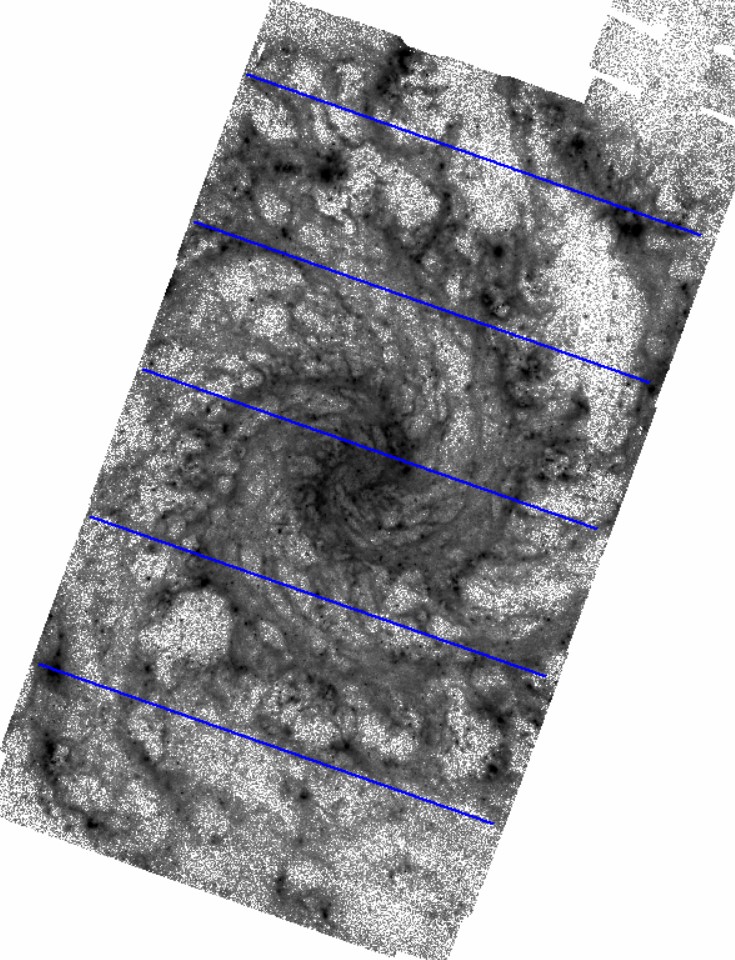}
    \end{center}
    \caption{NGC 628 at $10\mu$m with sample scan directions like those used to determine the PS, as in Fig.~\ref{fig:628image}.  The lines are separated by 500 pixels, which is $40^{\prime\prime}$ and 1.91 kpc.
    }
\label{fig:628_1000_image}
\end{figure}

\begin{figure}[H]
    \includegraphics[width=8.5cm]{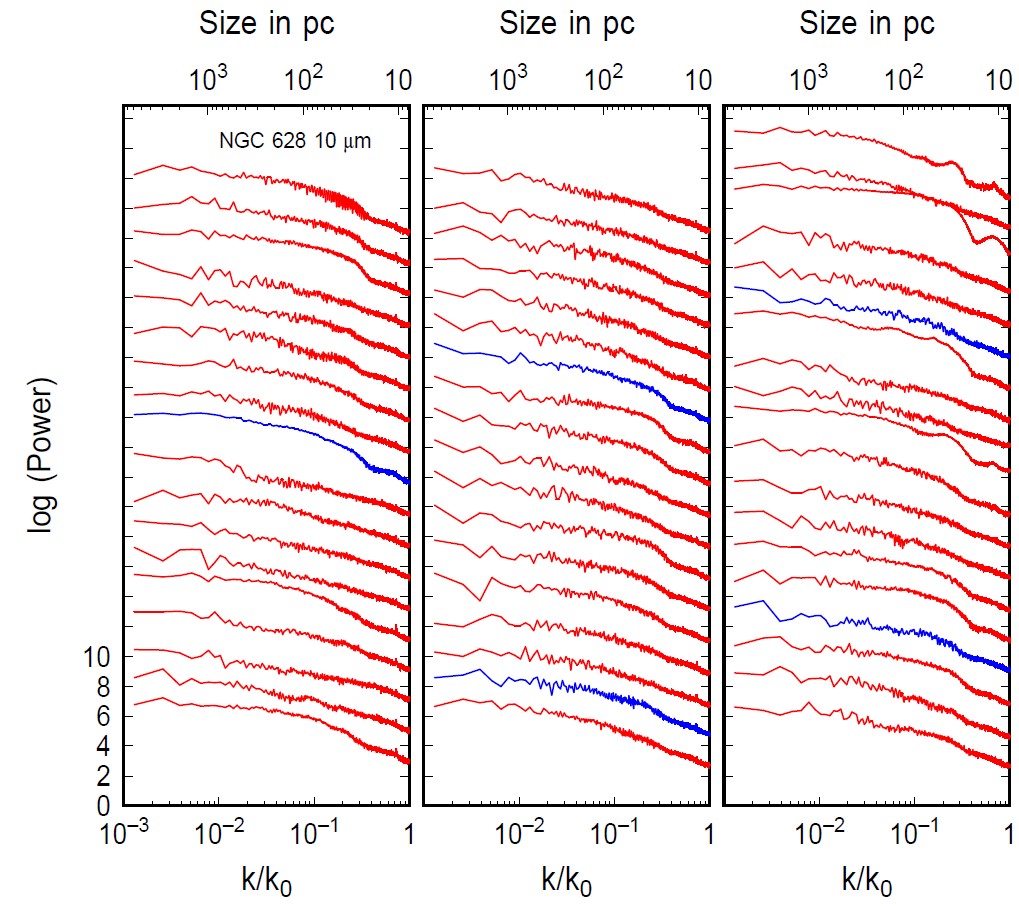}
    \caption{The same as Fig.~\ref{fig:ps628_560_all} but at $10\mu$m.}
    \label{fig:628_1000_all}
\end{figure}

\begin{figure}[H]
    \includegraphics[width=8.5cm]{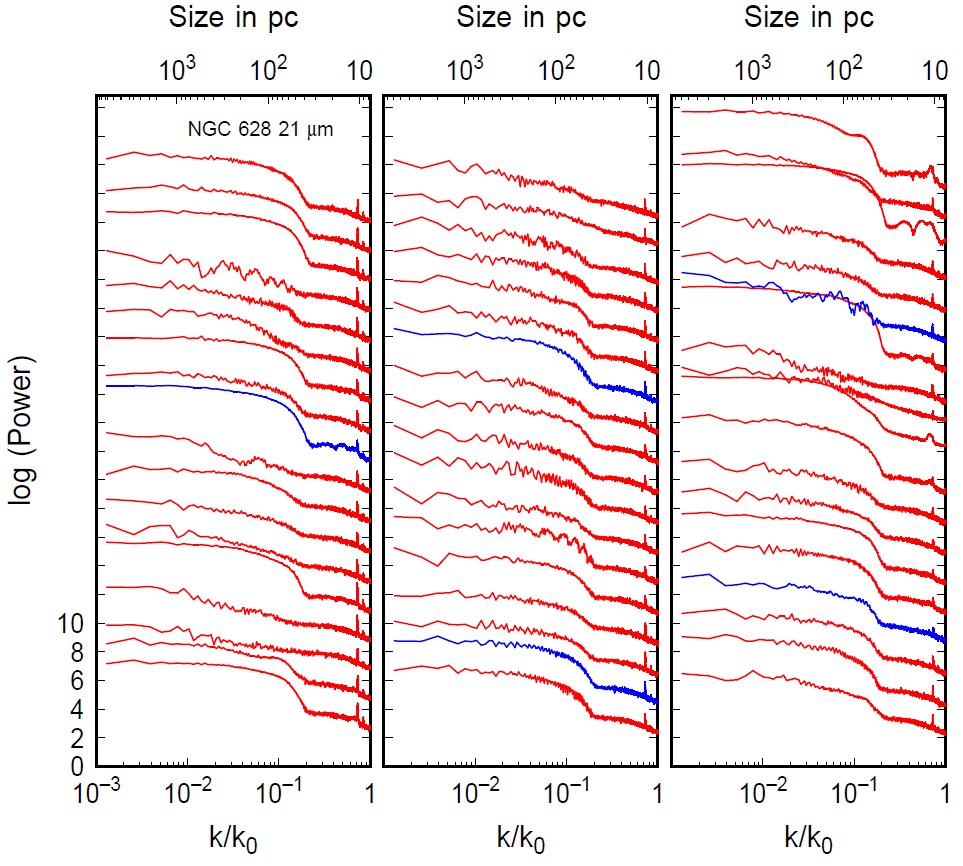}
    \caption{The same as Fig.~\ref{fig:ps628_560_all} but at $21\mu$m.}
    \label{fig:ps628_2100_all}
\end{figure}

\begin{figure}[H]
    \includegraphics[width=8.5cm]{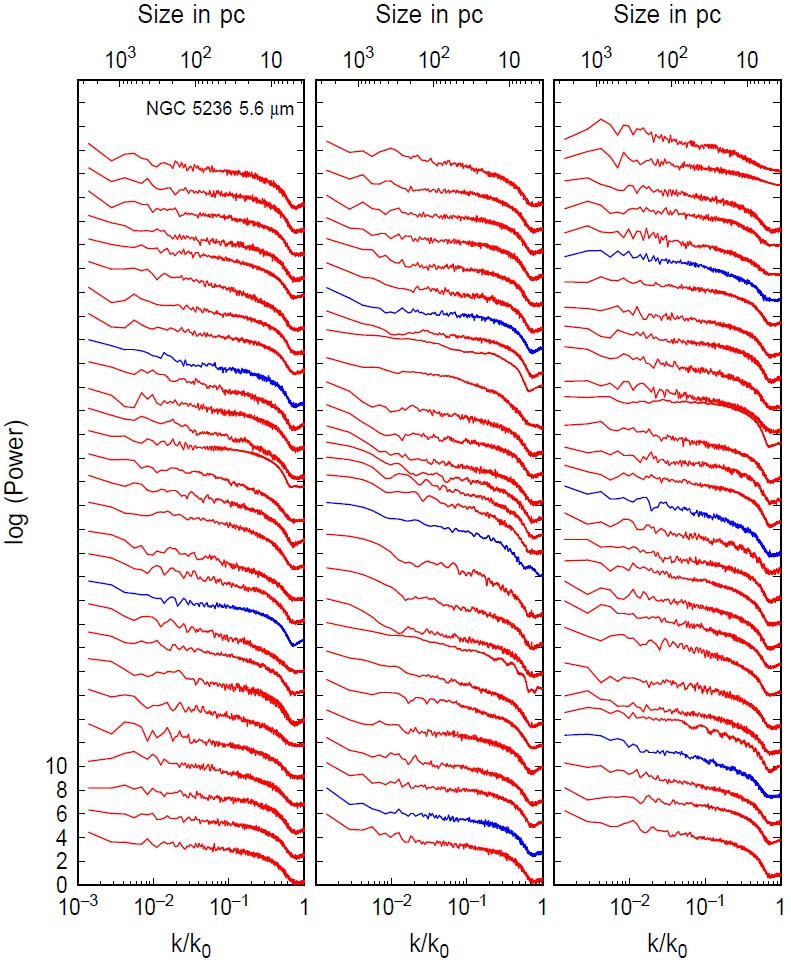}
    \caption{Averages of each 50 adjacent PS covering the NGC 5236 image in Fig.~\ref{fig:5236image}, shifted vertically for clarity. The 50-PS average corresponding to the southwest part of the image is at the bottom of the left-hand panel. Higher in the same panel, and then from bottom to top in the middle panel and bottom to top in the right-hand panel are the 50-PS averages in sequence from southwest to northeast in the image. }
    \label{fig:ps5236_560_all}
\end{figure}

\begin{figure}[H]
    \includegraphics[width=8.5cm]{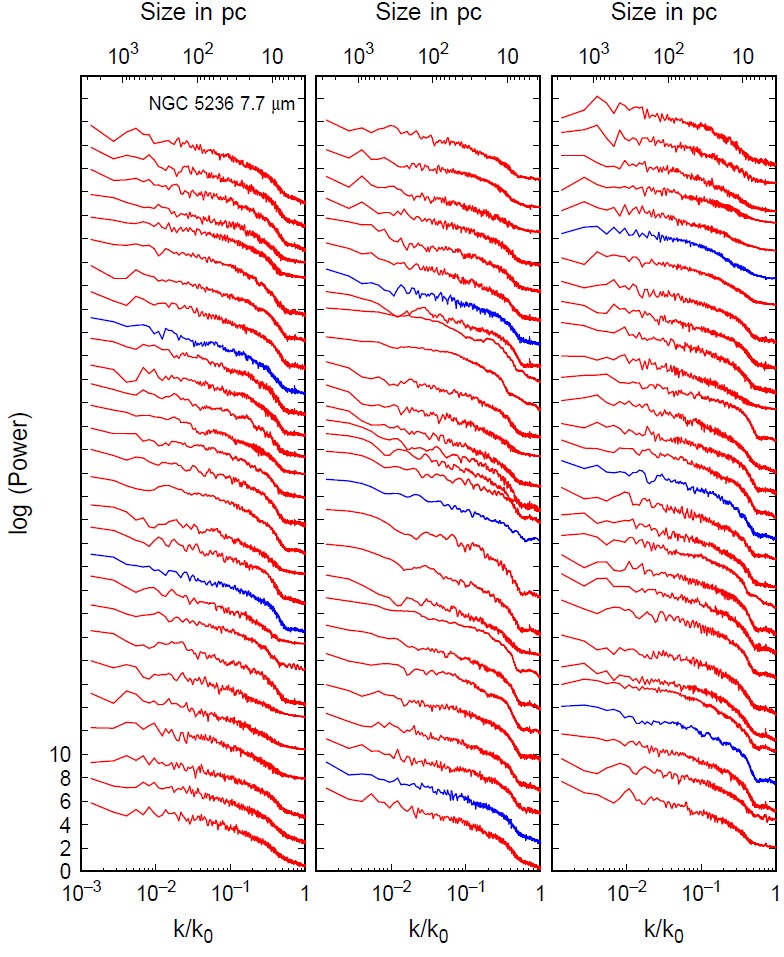}
    \caption{Same as Fig.~\ref{fig:ps5236_560_all} for NGC 5236 but at $7.7\mu$m.}
    \label{fig:ps5236_770_all}
\end{figure}

\begin{figure}[H]
    \includegraphics[width=8.5cm]{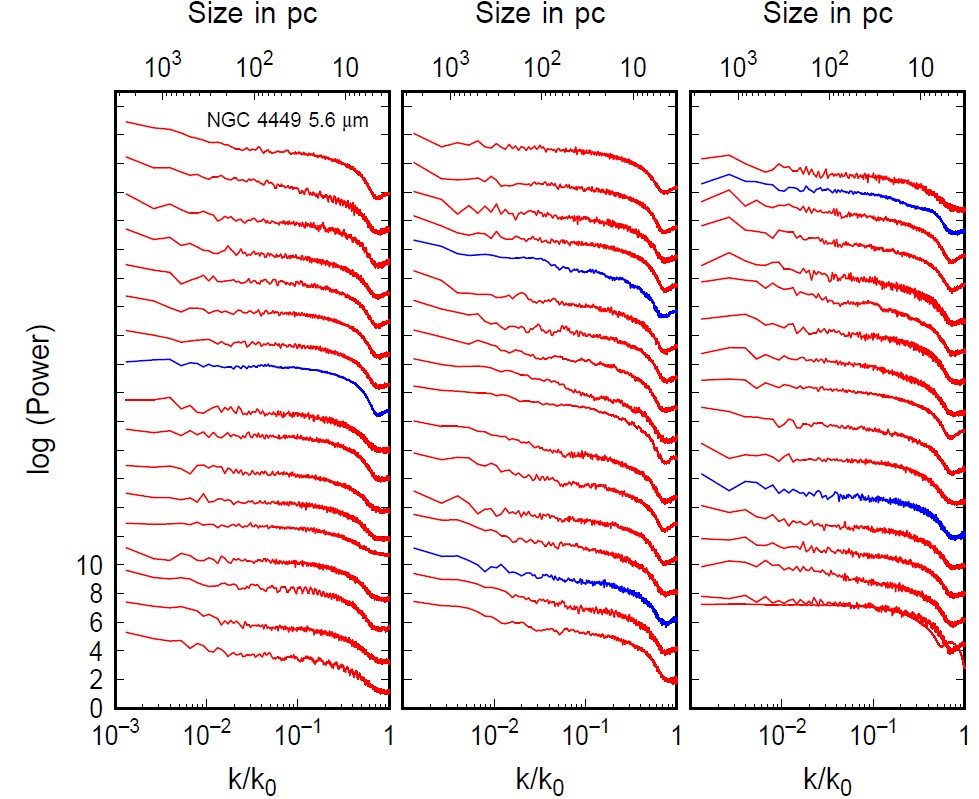}
    \caption{Averages of each 50 adjacent PS for the minor axis scans covering the $5.6\mu$m image in the left-hand panel of Fig.~\ref{fig:4449image}. Scans go from southwest in the lower part of the left-hand panel to northeast in the upper part of the right-hand panel.}
    \label{fig:ps4449_560_all}
\end{figure}

\begin{figure}[H]
    \includegraphics[width=8.5cm]{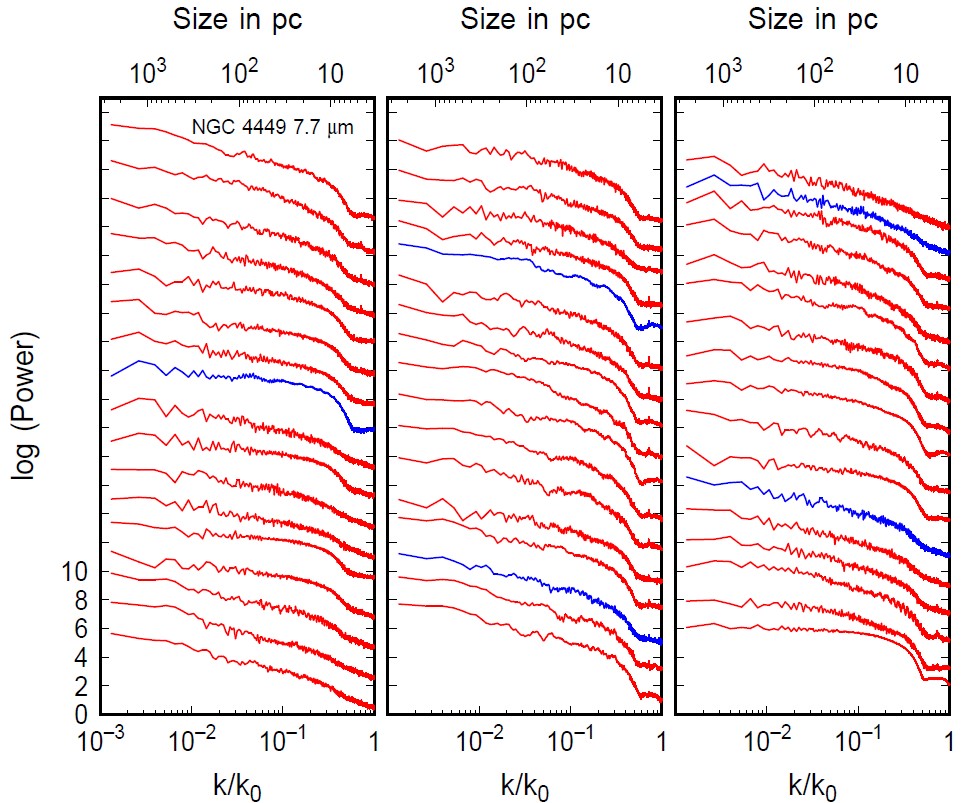}
    \caption{Same as Fig.~\ref{fig:ps4449_560_all} but for $7.7\mu$m. }
    \label{fig:ps4449_770_all}
\end{figure}

\begin{figure}[H]
    \begin{center}
        \includegraphics[width=6cm]{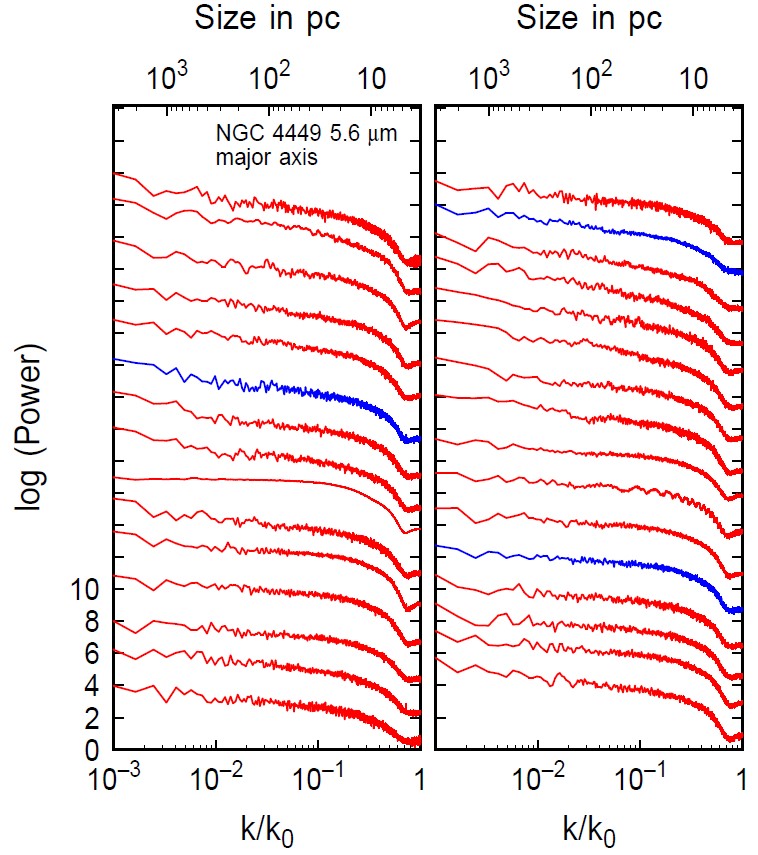}
    \end{center}
    \caption{Averages of each 50 adjacent PS for the major axis scans covering the $5.6\mu$m image in Fig.~\ref{fig:4449image}.}
    \label{fig:ps4449_560_major_all}
\end{figure}

\begin{figure}[H]
    \begin{center}
        \includegraphics[width=6cm]{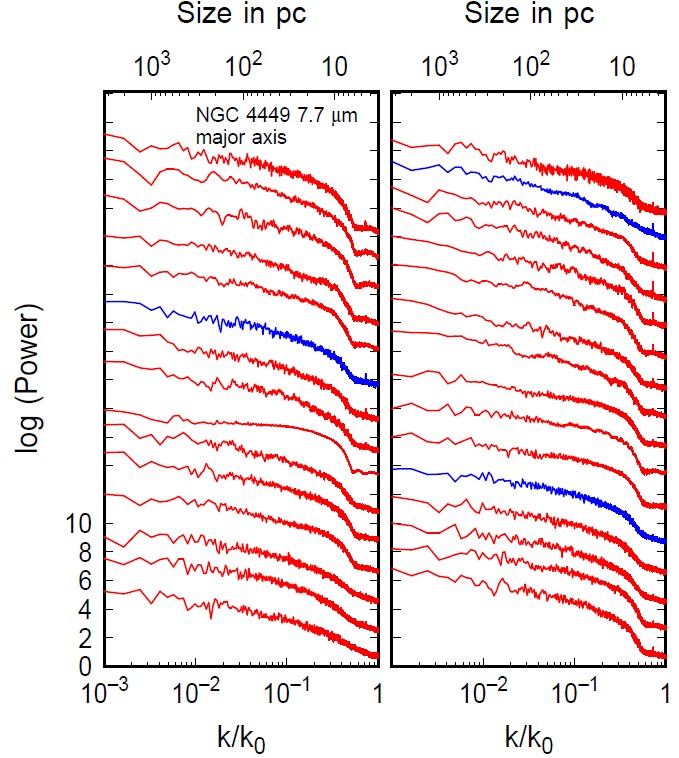}
    \end{center}
    \caption{Averages of each 50 adjacent PS for the major axis scans covering the $7.7\mu$m image in Fig.~\ref{fig:4449image}.}
    \label{fig:ps4449_770_major_all}
\end{figure}

\begin{figure}[H]
    \begin{center}    
        \includegraphics[width=6cm]{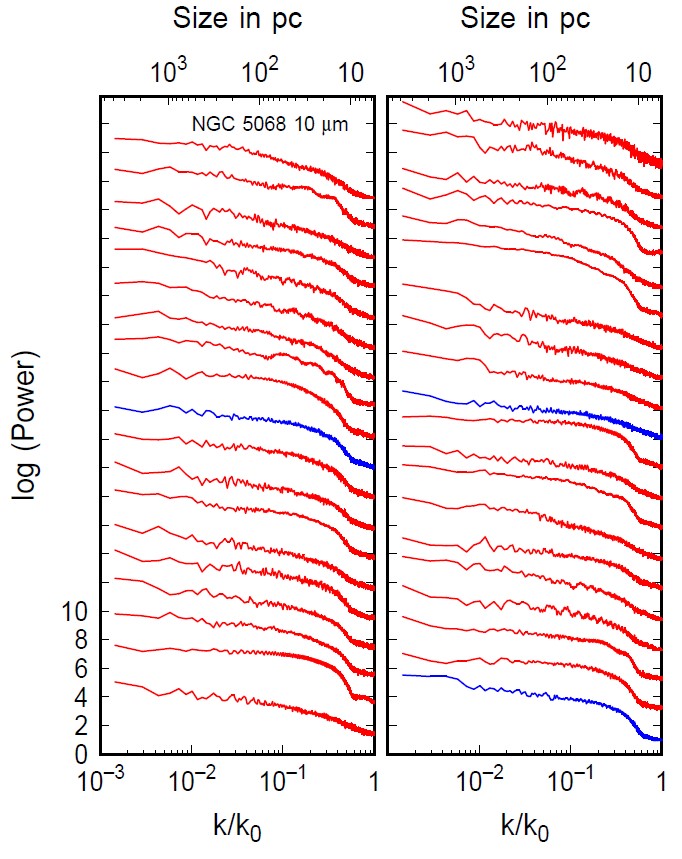}
    \end{center}
    \caption{Averages of each 50 adjacent PS for the $10\mu$m image of NGC 5068 in Fig.~\ref{fig:5068image}. Scans go from southeast in the lower part of the left-hand panel to northwest in the upper part of the right-hand panel.}
    \label{fig:ps5068_1000_all}
\end{figure}

\begin{figure}[H]
    \begin{center}    
        \includegraphics[width=6cm]{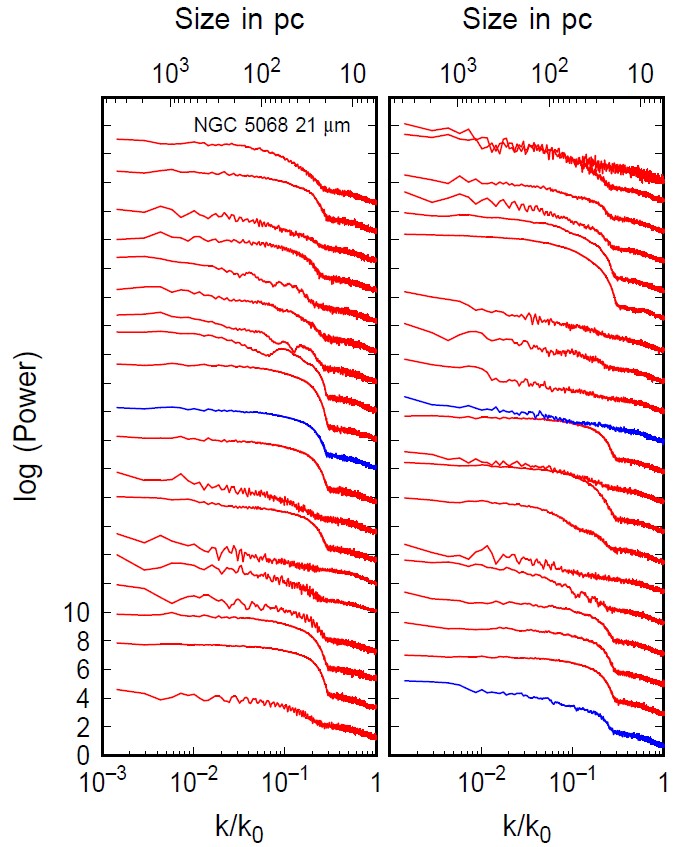}
    \end{center}
    \caption{Averages of each 50 adjacent PS for the $21\mu$m image of NGC 5068.}
    \label{fig:ps5068_2100_all}
\end{figure}

\begin{figure}[H]
    \begin{center}
        \includegraphics[width=6.5cm]{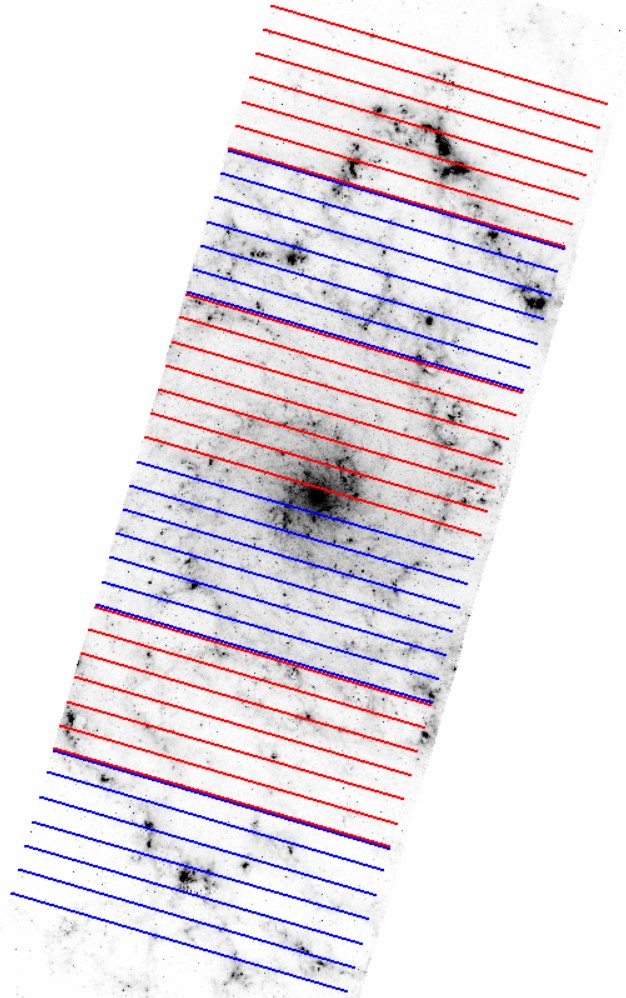}
    \end{center}
    \caption{NGC 628 at $5.6\mu$m with 6 regions having 600 scans each, indicated by the different colors, and in which $\sim30$ scans each were selected with the lowest intensity peaks for deriving the average PS shown in Fig.~\ref{fig:628_thresholds}. The lines are separated by 100 px, and there is one intensity scan per px.}
    \label{fig:628_560_minor_image}
\end{figure}

\begin{figure}[H]
    \begin{center}
        \includegraphics[width=6.5cm]{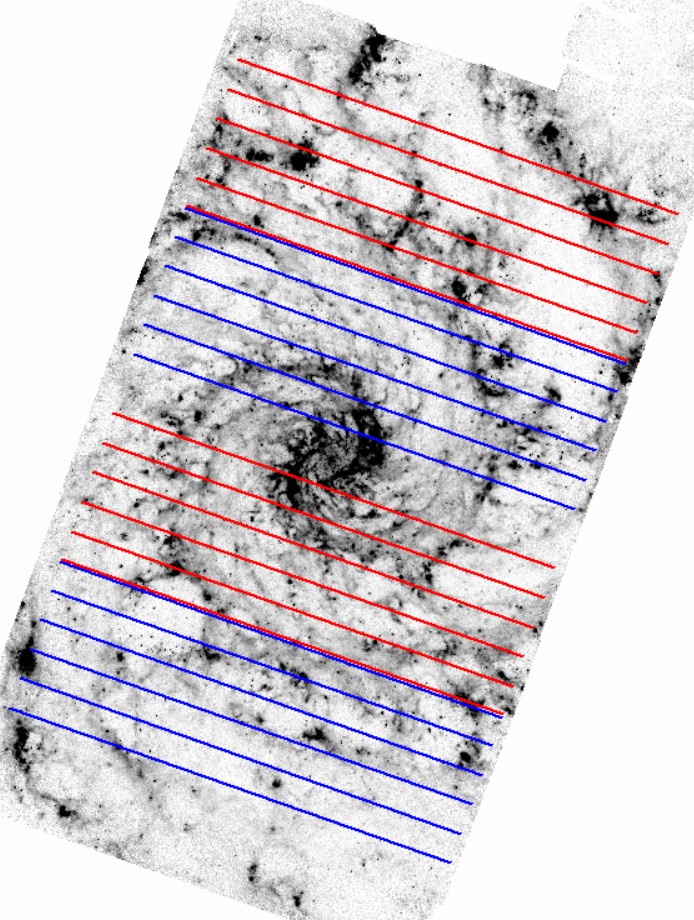}
    \end{center}
    \caption{NGC 628 at $10\mu$m with 4 regions having 500 scans each, indicated by the different colors, and in which $\sim30$ scans each were selected with the lowest intensity peaks for deriving the average PS shown in Fig.~\ref{fig:628_thresholds}. The lines are separated by 100 px, and there is one intensity scan per px.}
    \label{fig:628_1000_minor_image}
\end{figure}

\begin{figure}[H]
    \begin{center}
        \includegraphics[width=6.5cm]{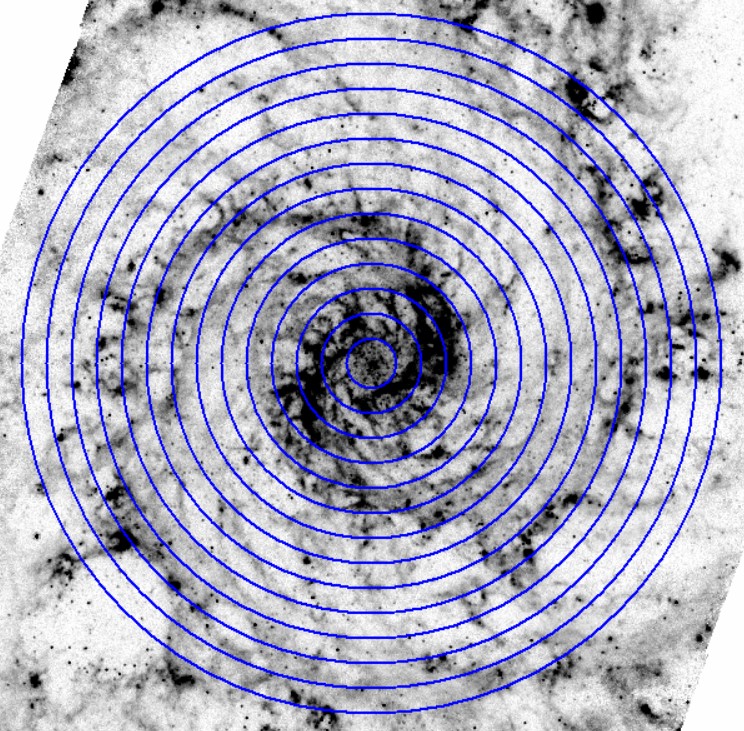}
    \end{center}
    \caption{NGC 628 at $10\mu$m with circles at the positions of azimuthal scans, separated by 50 px, which is 10 scans. We selected  $\sim30$ scans with the lowest intensity peaks from the total of 133 scans between radii of 100 px (20 scans) and 765 px (153 scans). The PS of these select scans were averaged together  to make the PS in Fig.~\ref{fig:628_PS_circles}. }
    \label{fig:628_1000_circles_image}
\end{figure}

\end{appendix}

\end{document}